\documentclass[12pt]{iopart}

\usepackage{amssymb}
\usepackage{iopams}

\usepackage[]{graphicx}
\usepackage[]{color}

\newcommand{\BR}{{\mathbb R}}
\newcommand{\BN}{{\mathbb N}}

\newcommand{\BC}{{\mathbb C}}

\newcommand{\cl}{C \kern -0.1em \ell}

\newcommand{\proof}{\bf {Proof:} \rm}
\newcommand{\qed}{$\blacksquare$}
\newtheorem{theorem}{Theorem}[section]
\newtheorem{remark}{Remark}[section]
\newtheorem{lemma}{Lemma}[section]
\newtheorem{proposition}{Proposition}[section]
\newtheorem{corollary}{Corollary}[section]

\begin{document}

\title{Fock spaces, Landau operators and the time-harmonic Maxwell equations}

\author{Denis Constales$^1$, Nelson Faustino$^2$ and Rolf S\"oren Krau\ss har$^3$}

\address{$^1$ Department of Mathematical Analysis,
Ghent University, Building S-22, Krijgslaan 281, B-9000 Ghent, Belgium; and Laboratory for Chemical Technology, Department of Chemical Engineering, Ghent University, Building S-5, Krijgslaan 281, B-9000 Ghent Belgium}

\address{$^2$ Centre for Mathematics, University of Coimbra, Largo D. Dinis, Apartado 3008, P-3001 -- 454 Coimbra, Portugal}

\address{$^3$ Fachbereich Mathematik,
Arbeitsgruppe Algebra, Geometrie und Funktionalanalysis,
Schlo{\ss}gartenstra{\ss}e 7, D-64289 Darmstadt}

\eads{\mailto{Denis.Constales@gmail.com},
\mailto{nelson@mat.uc.pt}, \mailto{krausshar@mathematik.tu-darmstadt.de}}

\begin{abstract}
We investigate the representations of the solutions to Maxwell's
equations based on the combination of hypercomplex
function-theoretical methods with quantum mechanical methods.~Our
approach provides us with a characterization for the solutions to
the  time-harmonic Maxwell system in terms of series expansions
involving spherical harmonics resp.\ spherical monogenics. Also, a
thorough investigation for the series representation of the
solutions in terms of eigenfunctions of Landau operators that
encode $n-$dimensional spinless electrons is given.

This new insight should  lead to important investigations in the
study of regularity and hypo-ellipticity of the solutions to
Schr\"odinger equations with natural applications in relativistic
quantum mechanics concerning massive spinor fields.
\end{abstract}




\maketitle

\section*{Dedicatory}

This paper is dedicated to Professor Richard Delanghe on the occasion of his seventieth birthday.

\section{Introduction}

\subsection{The Scope of Problems}

The problem of constructing exact solutions for the time-harmonic Maxwell equations plays an important role in investigations of relativistic particles (bosons and fermions) in electromagnetic fields in terms of relativistic quantum mechanics.

For a domain $\Omega \subset \BR^3$, and for an electromagnetic field with electric and magnetic components ${\bf E}:\Omega \rightarrow \BR^3$ and ${\bf H}:\Omega \rightarrow \BR^3$, respectively, the time-harmonic Maxwell equations with complex electric conductivity $\sigma:=\sigma^* - i \omega\varepsilon$, dielectric constant $\varepsilon$, magnetic permeability $\mu$ and medium electrical conductivity $\sigma^*$, are described in terms of the
following coupled system of equations:
\begin{eqnarray}
\label{TimeHarmonicMaxwell}\left\{\begin{array}{ccc}
{\rm rot}\; {\bf H} &=& \sigma {\bf E} \\  {\rm rot} \;{\bf E} &=& i
\omega \mu
{\bf H} \\
{\rm div}\; {\bf H} &=& 0 \\ {\rm div}\; {\bf E} &=& 0
\end{array}\right.
\end{eqnarray}

As is well known, the above system of equations may be formulated in terms of differential forms or integral representations. In particular, the electrical and
magnetic fields, ${\bf E}:\Omega \rightarrow \BR^3$ and ${\bf H}:\Omega \rightarrow \BR^3$ respectively, are also solutions of homogeneous
Helmholtz equations with respect to the square of the medium wave number $\lambda^2:= i \omega \mu \sigma^* + \omega^2 \mu \varepsilon
= i \omega \mu \sigma \in \mathbb{C}$:
\begin{eqnarray}
\label{HelmholtzEquation}\left\{\begin{array}{ccc}\Delta {\bf E} + \lambda^2 {\bf E} & = & 0\\
\Delta {\bf H} +\lambda^2 {\bf H} & = & 0
\end{array}\right.
\end{eqnarray}
For further details see \cite{Xu2,KS,Mbook}.

In the case where $\lambda$ is a pure imaginary value, say $\lambda=i\alpha$ ($\alpha \in \mathbb{R}$), it should be noticed that the equation~(\ref{HelmholtzEquation}) has also another important physical meaning. If we assign $\alpha = \frac{mc}{\hbar}$, where $m$ represents the mass of a particle, $c$ the speed of light and $\hbar$ the Planck constant, then this equation, which then is nothing else than the Klein-Gordon equation, correctly describes the spin-less pion. See also \cite{Kra09} for more details on this particular case.

The formulation of (\ref{TimeHarmonicMaxwell}) and (\ref{HelmholtzEquation}) (cf.\ \cite{KS}, Chapter 2)
in terms of the Dirac operator $D=-{\rm div}+{\rm rot}$, allows us to describe solutions of both systems in terms of displacements of $D$, say $D \mp \lambda I$, in the skew-field of quaternions.
Indeed, the vector fields ${\bf E}$ and ${\bf H}$ that satisfy (\ref{TimeHarmonicMaxwell}) coincide exactly with the solutions to the equation $(D - \lambda I)f=0$ in that domain.
In view of the factorization of the Helmholtz operator $\Delta+\lambda^2I$ in the form $\Delta+\lambda^2 I=-(D-\lambda I)(D+\lambda I)$, we can express the solutions ${\bf E}$ and ${\bf H}$ of (\ref{HelmholtzEquation}) in terms of the function ${\bf F}=(D+\lambda I)\left[{\bf E}+i {\bf H}\right]$. That function belongs to $\ker(D-\lambda I)$ and in turn allows us to re-express the electric and magnetic components for (\ref{TimeHarmonicMaxwell}).

An alternative geometric structure to describe (\ref{TimeHarmonicMaxwell}) and (\ref{HelmholtzEquation}) in terms of massive spinor fields is the setting of Clifford algebras (cf.\ \cite{KS}, Chapter 3).
Clifford algebras (see Section \ref{CliffordAlgebra})
endow finite dimensional quadratic vector spaces with an
additional multiplication operation which shall be understood in the language of differential forms as the combined action in terms of wedge and contraction operators (cf.\ \cite{GM92}, page 18).

When passing from the coordinate vector variable
$x=\sum_{j=1}^n x_j \e_j$ to polar coordinates:
\begin{eqnarray*}
\begin{array}{lll}
 x=r \theta &
\mbox{with} ~~r=|x| & \mbox{and}~~ \theta=\frac{1}{r}x,
\end{array}
\end{eqnarray*}
the normal derivative along the unit vector $\widehat{n}=\frac{x}{r}$ given by $\partial_{\widehat{n}}=r\frac{\partial}{\partial r}$ coincides with the so-called Euler operator $E$ (see Subsection \ref{ospClifford}) while $D$
is given by $$D=\theta\left( \frac{\partial}{\partial r}+ \frac{1}{r}
\Gamma\right).$$
Here, $\Gamma$ is the so-called Gamma operator or spherical Dirac operator (see Subsection \ref{ospClifford}).

The Laplace operator $\Delta$ corresponds to the decomposition that involves the action of the classical
Laplace-Beltrami operator
$\Delta_{LB}=((n-2)I-\Gamma)\Gamma$:
$$\Delta=-D^2=\frac{\partial^2}{\partial r^2}+\frac{n-1}{r}I+\frac{1}{r^2}\Delta_{LB}.$$

As explained in \cite{Xu1} and \cite{CCK06} the part of the solutions to the equation $(D - \lambda I)f=0$ that is regular inside a ball centered around the origin can explicitly be expressed in terms of finite sums of homogeneous polynomials from $\ker D$ multiplied with particular Bessel $J$-functions of integer resp.\ half-integer parameter. Outside the ball we get a similar representation, but the Bessel $J$ functions are then replaced by the corresponding Bessel $Y$ functions of the same parameter.

Following $\cite{PV,LR}$ and others, the spherical analogue of the operator $D-\lambda I$ is the operator $\Gamma-\lambda I$.
Its kernel consists of the solutions to the spherical time-harmonic Maxwell operator and can be written in terms of sums over standard hypergeometric functions $_2F_1$ multiplied with homogeneous polynomials in $\ker D$.
 Furthermore, in the recent work \cite{CCK09} it has been shown that the solutions to the time harmonic Maxwell system on a sphere of radius $R > 0$ coincides with the null solutions to the radial type operator  $D-\lambda I-\frac{1}{R}E$. The case described in the earlier works \cite{PV,LR} arises as particular subcase. When $R \to + \infty$, the system $(D-\lambda I-\frac{1}{R}E)f=0$ simplifies to $(D - \lambda I)f=0$ where we are dealing with the null solutions to the time-harmonic Maxwell operator in the Euclidean flat space.

 Apart from the development of function theoretical tools to compute analytically the solutions to the time-harmonic Maxwell equations (\ref{TimeHarmonicMaxwell}) and the homogeneous Helmholtz equations (\ref{HelmholtzEquation}), there is also considerable interest in the study of the spectra of Landau operators from the border view of physics and mathematics.

 While the study of Landau operators has its roots in the construction of coherent states for relativistic Klein-Gordon and Dirac equations (see \cite{BBG76} and references given there), the surge of interest in these operators arose in the study of the possible occurrence of orbital electromagnetism (cf.~\cite{GHJ02}) as well as in the study of pseudo-differential operators on modulation spaces (cf.~\cite{DeGossonLuef09}) and quantum representations of Gabor-windowed Fourier analysis (cf.~\cite{BrackenWatson10}).

  An extended survey of relativistic quantum mechanics and the construction of coherent states can be found in \cite{Folland08} (see Chapter 4) and \cite{Perelomov86} (see Chapter 19), respectively; for a deep understanding of the link between Gabor-windowed Fourier analysis and the Heisenberg-Weyl group or of the Weyl transform with Hermite series expansions we may refer to \cite{Thangavelu93}.
Recently, in \cite{ZKI10} the authors proposed a meaningful formulation of Landau operators in the quaternionic field that yields from the sub-Laplacian that arises in the quaternionic version of the Weyl-Heisenberg group.

 From the border view of Clifford algebras, the extension of the above construction that describes the topological laws underlying the time-harmonic Maxwell equations (\ref{TimeHarmonicMaxwell}) shall be defined as follows:

 Let us consider the following Hamiltonian operator with mass $m$, frequency $\omega$ and potential energy
 $\frac{m \omega^2}{2}\left|x\right|^2$ that encodes the Helmholtz operator $\Delta+\lambda^2I$:
\begin{eqnarray*}
\label{HamiltonianHelmholtz} \tilde{\mathcal{H}}_{\lambda}=-\frac{1}{2m}(\Delta+\lambda^2I) +\frac{m \omega^2}{2}\left|x\right|^2I.
\end{eqnarray*}

We define the Landau operator $\mathcal{H}_\lambda$ as the superposition of the Hamiltonian operator $\tilde{\mathcal{H}}_{\lambda}$ by a symmetric gauge term $\mathcal{L}_\lambda$ which involves the spherical Dirac operator $\Gamma$:
\begin{eqnarray}
\label{LandauOperator}\mathcal{H}_\lambda=\tilde{\mathcal{H}}_{\lambda}+\frac{m\omega^2}{2}\mathcal{L}_\lambda.
\end{eqnarray}

Along this paper $\mathcal{L}_\lambda$ corresponds to the special choice  $$\mathcal{L}_\lambda=-\frac{2\lambda}{n}\left(xI-2\lambda\left(I-\frac{1}{n}\Gamma\right)\Gamma\right)$$ that shall be understood as a certain sort of electromagnetic counterpart for the Laplace-Betrami operator $\Delta_{LB}$ defined above.
Here we would like to stress that when restricted to dimension $2$ or $4$, the spherical Dirac operator $\Gamma$ is equivalent to the orbital angular momentum operators that appear in \cite{GHJ02,ZKI10}.

\subsection{Motivation and Main Results}

The theory of Segal-Bargmann spaces introduced independently by Segal (cf.\ \cite{Segal63}) and Bargmann (cf.\ \cite{Bargmann61}) and popularized by Newman and Shapiro (cf.~\cite{NewmanShapiro66}) has its deep roots in the Fock space formalism (cf.\ \cite{Fock32}) and states that the spaces of analytic functions isomorphic to $L_2(\BC^n)$ may be realized as a canonical isomorphism in terms of the Weyl-Heisenberg algebra $\mathfrak{h}_n$.

Later with the approach of Howe (cf.\ \cite{Howe80}) it was shown that Segal-Bargmann spaces are linked with the Schr\"odinger representation of the Heisenberg group $\mathbb{H}^n$.
A full development of those ideas in the context of harmonic analysis and the theory of special functions can be found
in the books of Folland (cf.~\cite{Folland89}) and Thangavelu (cf.~\cite{Thangavelu93}), respectively; connections with the theory of coherent states can be found in the book of Perelomov (cf.~\cite{Perelomov86}) and
in the paper of W\"unsche (cf.\ \cite{Wunsche91}); in the context of Clifford analysis the study of representations for the
Heisenberg group was initiated by Kisil
in \cite{Kisil93} and lately generalized by Cnops and Kisil (cf.~\cite{CnopsKisil99}) for nilpotent Lie groups.

Also, the generalization of the theory of Segal-Bargmann spaces during the last decade by Askour, Intissar and
Mouyan (cf.\ \cite{AIM00}) and Vasilevski (cf.~\cite{Vasilev00}) led recently to
ubiquitous developments in the theory of Gabor-Window
Fourier analysis \cite{AbreuACHA10,AbreuMon10,BrackenWatson10}. From the border view of the Heisenberg group, it was recently shown
in \cite{DeGossonLuef09} that the Schr\"odinger representation and alike play an important role in the study of pseudo-differential operators establishing structural properties between the Weyl calculus and the Landau-Weyl calculus.

One of the goals of this paper is to emphasize the great potential of the quantum mechanical formalism in Clifford analysis for the study of the solutions of the time-harmonic Maxwell equations (\ref{TimeHarmonicMaxwell}), namely the Fock representation of (poly-)monogenic functions and the construction of the eigenspaces for the Hamiltonian operator
\begin{eqnarray}
\label{Hamiltonian} \mathcal{H}=-\frac{1}{2m}\Delta +\frac{m \omega^2}{2}|x|^2I
\end{eqnarray}
as dense linear subspaces of the Clifford-valued Schwartz space $\mathcal{S}(\BR^n;\BR_{0,n})$ that encode the symmetries of $\mathfrak{osp}(1|2)$.~When restricted to the algebra of Clifford-valued polynomials $\mathcal{P}$, these spaces
have turned out to be an adequate setting for the construction of series representations in terms of Clifford-Hermite functions or polynomials, respectively.

Although Clifford-Hermite functions resp.\ polynomials expansions and their applications are well-known in harmonic analysis, Fourier analysis and wavelet analysis
(cf.\ \cite{Sommen88,BS00,DBieSommen07,DBieXu10}) they have not received that much attention in quantum mechanics after the approach of Cnops and Kisil \cite{CnopsKisil99} in which the representation of nilpotent Lie groups like the Heisenberg group $\mathbb{H}^n$ arises in the hypercomplex formulation.

Let us get some motivation from the representation of nilpotent Lie groups (see \cite{Perelomov86}, Chapter 10).
If one considers operators of the type $\exp\left( \frac{\lambda}{n}(D-X)\right)$ as the Clifford extension of the displacement/Heisenberg-Weyl operators (cf.\ \cite{Wunsche91,DeGossonLuef09}) to the Heisenberg group $\mathbb{H}^n$, we will get an intriguing link that allows us to describe the solutions of the PDE system
$$ \left(D-\lambda I+\frac{2\lambda}{n}\Gamma\right)f_\lambda=g_\lambda,~~\mbox{with}~g_\lambda \in \ker\left(D-\lambda I+\frac{2\lambda}{n}\Gamma\right)^s$$
as a generating-type function involving Clifford-Hermite polynomials.

The most interesting output of this paper is the link between the Landau operator $\mathcal{H}_\lambda$, the standard Hamiltonian operator $\mathcal{H}$ that follows from the covariant action $xI \mapsto xI-\lambda I+\frac{2\lambda}{n}\Gamma$ on the spherical potential $\frac{m \omega^2}{2}\left|x\right|^2I$ and the (apparently) new connection between the series expansions of $D-\lambda I+\frac{2\lambda}{n}\Gamma$ and the eigenfunctions for the Landau operator $\mathcal{H}_\lambda$.

\subsection{Organization of the paper}

In this paper we have tried to present a self-contained exposition in a concise style leaving the proof of the technical lemmata to \ref{HarmonicOscillatorAppendix} and \ref{MaxwellEquationsAppendix}. For the sake of easy readability of the paper, we add in \ref{Table} a table with the main symbols used along the main body of the paper.

The outline is as follows: In Section \ref{CliffordSetting} we will start to recollect some basic features of Clifford algebras and Clifford analysis; we may refer for example to \cite{GM92,DSS92} in which an extended survey is presented.

Section \ref{HarmonicOscillator} will be devoted to the study of the spectra of the Hamiltonian (\ref{Hamiltonian}) in terms of Clifford algebra-valued functions.
We will start to review some basic facts regarding Hermite functions in terms of Weyl-Heisenberg symmetries (cf.\ \cite{Folland89,Thangavelu93,Folland08}). Afterwards, using the $\mathfrak{osp}(1|2)$ symmetries encoded in the Clifford-valued operators, we will get a full descriptiona for the eigenspaces of (\ref{Hamiltonian}) in terms of $\mathcal{H}_0=\frac{1}{2}(-\Delta+|x|^2I)$ by means of $\mathfrak{osp}(1|2)$ symmetries that comprise the function spaces spanned by Clifford-Hermite polynomials and Clifford-Hermite functions giving in this way a quantum mechanical interpretation for the results obtained in \cite{Sommen88}.

Finally, in Section \ref{MaxwellEquations} we will study the structure of the solutions of the time-harmonic Maxwell equations (\ref{TimeHarmonicMaxwell}) subjected to an orbital angular momentum action encoded in the spherical Dirac operator $\Gamma$
and its interplay with the theory of spherical monogenics and with the eigenfunctions underlying the Landau operator $\mathcal{H}_\lambda=\tilde{\mathcal{H}}_\lambda+\mathcal{L}_\lambda$ (see equation (\ref{LandauOperator})).

\section{The Clifford analysis setting}\label{CliffordSetting}

\subsection{Clifford Algebras}\label{CliffordAlgebra}

Let $\BR^n$ be endowed with the non-degenerate bilinear symmetric form
$\mathcal{B}(\cdot,\cdot)$ of signature $(p,q)$, with $p+q=n$ and let $\e_1,\e_2,\ldots,\e_n$ be an orthogonal basis of $\BR^n$
satisfying
\begin{center}
    $\begin{array}{ccll}
        \mathcal{B}(\e_j,\e_j) & = & -1, & j=1,\ldots,p \\
        \mathcal{B}(\e_j,\e_j) & = & 1,  & j=p+1,\ldots,n \\
        \mathcal{B}(\e_j,\e_k) & = & 0,  & j \neq k.
    \end{array}$
\end{center}

We denote by $\BR_{p,q}$ the real Clifford algebra of signature $(p,q)$ generated by the
identity $1$ and the standard basis elements $\e_j$ modulo the relations
\begin{equation}
\label{ejek}    \{ \e_j,\e_k\}=  2 \mathcal{B}(\e_j,\e_k).
\end{equation}
Here $\{ {\bf a},{\bf b}\}:={\bf a}{\bf b}+{\bf b}{\bf a}$ denotes the anti-commutator between ${\bf a}$ and ${\bf b}$.

The elements of $\BR_{p,q}$ are called Clifford numbers. The relation (\ref{ejek}) is the so-called Kronecker factorization.
For $\underline{\e}=(\e_1,\e_2,\ldots,\e_n)$ and for each $\alpha \in \{ 0,1\}^n$, we set $\underline{\e}^\alpha=\e_1^{\alpha_1}\e_2^{\alpha_2}\ldots \e_n^{\alpha_n}$ as the canonical basis for $\BR_{p,q}$. For $\alpha=\underline{0}$, we put $\underline{\e}^{\underline{0}}=1$.

An element ${\bf a} \in \BR_{p,q}$ is called an $r-$vector if ${\bf
a}$ may be written as a sum of elements of the form $a_\alpha \underline{e}^\alpha$,~with $|\alpha|=r$ (i.e. $\alpha$ has $r$ non-vanishing indices). The space of all $r-$vectors
is denoted by $\BR^r_{p,q}$ and $[\cdot]_r: \BR_{p,q}
\rightarrow \BR^r_{p,q}$ stands for the projection operator of
$\BR_{p,q}$ onto $\BR^r_{p,q}$ defined as $[{\bf a}]_r=\sum_{|\alpha|=r}a_\alpha\underline{\e}^\alpha.$
This leads to the identification of $\BR$ with the subspace $\BR_{p,q}^0$ (scalars), $\BR^n$
with $\BR_{p,q}^1$, (Clifford vectors of signature $(p,q)$)
and of the space of volume-forms with the subspace
$\BR_{p,q}^n$ generated by the single $n-$vector $\e_1 \ldots
\e_n$ (the so-called pseudoscalar).
Moreover, every element ${\bf a} \in \BR_{p,q}$ may be decomposed in a unique way as a finite sum of the form
${\bf a}=\sum_{r=0}^n [{\bf a}]_r$ and hence
$$\BR_{p,q}=\sum_{r=0}^n \oplus \BR_{p,q}^r.$$

We would like to stress that $\BR_{p,q}$ is in fact an algebra of radial type $R(\mathcal{S})$ generated by
$\mathcal{S}=\BR_{p,q}^1$:
\begin{eqnarray}
\label{radial}
\begin{array}{lll}
[\{x,y\},z]=0 & \mbox{for any}~{x,y,z \in \mathcal{S}}.
\end{array}
\end{eqnarray}
This means that there actually is no a priori defined linear space to which the vector variables $x=\sum_{j=1}^n x_j\e_j \in \mathcal{S}$ belong. Nevertheless, by only using (\ref{radial}) one can already deduce many properties but we shall not explore them here. Further details can be found in \cite{Sommen97} and  additionally in ~\cite{GM92} on page 17 to get the characterization of $\BR_{p,q}$ in terms of $R(\mathcal{S})$ modulo the center of $\BR_{p,q}$.

For $x \in \BR_{p,q}^1$ and ${\bf a}^r \in \BR_{p,q}^r$, the \textit{inner} product and the \textit{wedge} product on $\BR_{p,q}$ are  defined by
\begin{eqnarray}
\label{geometric_product2}\begin{array}{ccc}
x \bullet {\bf a}^r=[x  {\bf a}^r]_{r-1}=\frac{1}{2}\left( x{\bf a}^r-(-1)^r {\bf a}^r x \right) \\
x \wedge {\bf a}^r=[x {\bf a}^r]_{r+1}=\frac{1}{2}\left( x{\bf a}^r+(-1)^r {\bf a}^r x \right).
\end{array}
\end{eqnarray}
The geometric Clifford product is given by $x {\bf a}^r=x \bullet {\bf a}^r+x \wedge {\bf a}^r$.
Geometrically speaking, $x \in \BR^{1}_{p,q}$ is orthogonal (respectively, parallel) to
$y \in \BR^{1}_{p,q}$ if $ x \bullet y =0$ (respectively, $x \wedge y =0$).
 Therefore, the orthogonality between $x$ and $y$ leads to
$xy=-yx$ while the commutativity between $x$ and $y$ (i.e. $xy=yx$) occurs if $x$ is parallel to $y$.
 By (\ref{geometric_product2}), $x^2= x \bullet x$ is a
real number for any $x$. Hence, $x$ is invertible if and only if $x^2 \neq 0.$ In
this case the inverse $x^{-1}$ is given by $x^{-1}=\frac{x}{x^2}$ with $x^2=-\sum_{j=1}^p x_j^2+\sum_{j=p+1}^{p+q} x_j^2$. If $x^2 =0$ then either
$x$ is zero or it is a zero divisor and hence not invertible.

There are essentially two linear anti-automorphisms (reversion and conjugation)
and a linear automorphism (main involution) acting on $\BR_{p,q}$.
\begin{description}
    \item[\quad \,-] The \textit{main involution} is defined by
                     $$ \e_j` = -\e_j, \quad 1`=1, \quad (j=1,\ldots,n), \quad
                     (ab)`=a` b`, \,\,\,\,\forall \, a,b \in \BR_{p,q} .$$
    \item[\quad \,-] the \textit{reversion} is defined by
                     $$ \e_j^* = \e_j, \quad 1^*=1, \quad (j=1,\ldots,n), \quad
                     (ab)^*=b^* a^*, \,\,\,\,\forall \, a,b \in \BR_{p,q} .$$
    \item[\quad \,-] the \textit{conjugation} is defined by
                     $$ \e_j^\dag = -\e_j, \quad {1}^\dag={1}, \quad (j=1,\ldots,n),
                     \quad (ab)^\dag= b^\dag a^\dag, \,\,\,\,\forall \, a,b \in \BR_{p,q}.$$
\end{description}

We stress that {\it conjugation} can be obtained as a composition between {\it main involution} and {\it reversion} i.e.  $x^\dag = (x')^* = (x^*)', \,
\forall x \in \BR_{p,q}.$ From the definition we can derive the
action on the basis elements $\underline{\e}^\alpha$ by the rules:
$$
    (\underline{\e}^\alpha)` = (-1)^{|\alpha|} \underline{\e}^\alpha, \qquad (\underline{\e}^\alpha)^* = (-1)^{\frac{|\alpha|(|\alpha|-1)}{2}} \underline{\e}^\alpha,
    \qquad (\underline{\e}^\alpha)^\dag = (-1)^\frac{|\alpha|(|\alpha|+1)}{2} \underline{\e}^\alpha.
$$
In particular, if $x$ is a vector, then we obtain $x^\dag = x` = -x$ and
$x^*=x.$

The $\dag-$conjugation leads to the \index{Clifford inner product}Clifford algebra-valued inner product and its associated norm on $\BR_{p,q}$ given by $({\bf a},{\bf b})=[{\bf a}^\dag{\bf b}]_0$ and $|{\bf a}|^2=({\bf a},{\bf a}),$ respectively.
Notice that when ${\bf a}$ and ${\bf b}$ belong to $\BR_{p,q}^1$, their inner product and associated norm reduces to the classical inner product and norm on the ambient space $\BR^n$, respectively.

\subsection{Orthosymplectic Lie algebra representation of Clifford-valued operators}\label{ospClifford}

Let us restrict ourselves to the real Clifford algebra of signature $(n,n)$, $\BR_{n,n}$, in particular in its realization as the \index{algebra of endomorphisms}algebra of endomorphisms $\mbox{End}(\BR_{0,n}).$
Here and elsewhere, we will consider Clifford-valued functions on $\BR_{0,n}$, i.e., functions which can be decomposed in terms of $\BR_{0,n}^r-$valued functions:
\begin{eqnarray*}
\begin{array}{lll}
 f(x)=\sum_{j=0}^r [f(x)]_r, & \mbox{with}~~[f(x)]_r=\sum_{|\alpha|=r} f_\alpha(x){\bf e}^\alpha.
\end{array}
\end{eqnarray*}

Let us observe that, from (\ref{geometric_product2}) for any ${\bf a} \in \BR_{0,n}$, the {\it inner} product and the {\it wedge} product, $\e_j \bullet {\bf a}$ and $\e_j \wedge {\bf a}$, respectively, correspond to
\begin{eqnarray}
\label{geometric_product3}
\begin{array}{lll}
\e_j \bullet {\bf a}=\frac{1}{2}\left( \e_j{\bf a}- {\bf a}^* \e_j \right), &
\e_j \wedge {\bf a}=\frac{1}{2}\left( \e_j{\bf a}+{\bf a}^* \e_j \right),
\end{array}
\end{eqnarray}
where $*$ is the main involution defined in the preceding section.
This suggests to take the basic endomorphisms $\xi_j: {\bf a} \mapsto \e_j {\bf a}$ and $\xi_{j+n}: F(\underline{x}) \mapsto  {\bf a}^*\e_j$ acting on $\BR_{0,n}.$

It is clear that $\xi_j$ and $\xi_{j+n}$ correspond one-to-one to the generators of the algebra $\BR_{n,n}$ since $\xi_j(\xi_j {\bf a})=\e_j^2{\bf a}=-{\bf a}$, $\xi_{j+n}(\xi_{j+n} {\bf a})=({\bf a}^*\e_j)^*\e_j={\bf a}\e_j^*\e_j={\bf a}$ and
\begin{eqnarray*}
\begin{array}{lll}
\xi_{j}(\xi_{k}{\bf a})+\xi_{k}(\xi_{j}{\bf a})=0, & \mbox{for}~~j,k=1,\ldots,2n~~\mbox{with}~j\neq k.
\end{array}
\end{eqnarray*}
Moreover, the operator
actions $\e_j \wedge (\cdot) =\frac{1}{2}\left( \xi_{j}-\xi_{j+n} \right)$ and $\e_j \bullet (\cdot)=\frac{1}{2}\left( \xi_{j}+\xi_{j+n}\right)$ naturally give rise to a new basis for $\mbox{End}(\BR_{0,n})$, the so-called Witt basis for $\BR_{n,n}$ satisfying the relations below:
\begin{eqnarray*}
\begin{array}{llll}
\mbox{Grassmann identities:} & \left\{ \e_j \wedge (\cdot),\e_k \wedge (\cdot) \right\}=0=\left\{ \e_j \bullet (\cdot),\e_k \bullet (\cdot) \right\},   \\
\mbox{duality identities:} &\left\{ \e_j \bullet (\cdot),\e_k \wedge (\cdot) \right\}=\delta_{jk}I.
\end{array}
\end{eqnarray*}
where $I$ stands for the identity operator.

Next, we set $\mathcal{A}_n$ to be the algebra of differential operators given in terms of left endomorphisms $X_j:f(x) \mapsto x_jf(x)$, $\partial_{X_j}:f(x)\mapsto \frac{\partial f}{\partial x_j}(x)$ and $\xi_j:f(x) \mapsto \e_j f(x)$:
$$ \mathcal{A}_n=\mbox{span}\{ X_j,\partial_{X_j}, \xi_{j}~:~j=1,2,\ldots,n \}.$$
It is straightforward to see that $X_1,X_2,\ldots,X_j,\partial_{X_1},\partial_{X_2},\ldots,\partial_{X_n},I$ are the generators of the $(2n+1)-$dimensional Heisenberg-Weyl Lie algebra $\mathfrak{h}_n$:
\begin{eqnarray}
\label{WeylHeisenberg}
\begin{array}{lll}
[X_j,X_k]=0,& [\partial_{X_j},\partial_{X_k}]=0, & [\partial_{X_j},X_k]=\delta_{jk}I.
\end{array}
\end{eqnarray}
This is a simple consequence of the Leibniz rule underlying $\frac{\partial}{\partial x_j}$ and the mutual commutativity between the coordinates $x_j$ and partial derivatives $\frac{\partial}{\partial x_j}$.

We define the Dirac operator $D$ and the vector multiplication operator $X$ on $\BR_{0,n}$ as the following left endomorphisms:
\begin{eqnarray*}
\begin{array}{lll}
D=\sum_{j=1}^n \xi_j \partial_{X_j}, &X=\sum_{j=1}^n \xi_j X_j.
\end{array}
\end{eqnarray*}

Clearly, these operators are elements of $\mathcal{A}_n$. In terms of $D$ and $X$, $\mathcal{A}_n$ is equivalent to
$ \mathcal{A}_n=\mbox{span}\{ X,D, \xi_{j}~:~j=1,2,\ldots,n \}.$
Moreover, the Laplace operator $\Delta$ and the square of $X$ are also elements of $\mathcal{A}_n$ since $\Delta=-D^2$ and $X^2=-|x|^2I$.
Using the wedge product $\wedge$ and the dot product $\bullet$ introduced above, we further introduce the so-called Euler and Gamma operator as follows:
\begin{eqnarray}
\begin{array}{lllll}
E&=&X \bullet D&=&\sum_{j=1}^n X_j \partial_{X_j} \\
\Gamma&=&-X \wedge D &=& -\sum_{k=1}^n\sum_{j<k} \xi_j \xi_k (X_j \partial_{X_k}-X_k\partial_{X_j}).
\end{array}
\end{eqnarray}
The following lemma, in which we consider combinations between $X,D,E,\Delta$ and $\Gamma$, will be important in the sequel. It establishes that $E$ and $\Gamma$ are also elements of $\mathcal{A}_n$.
\begin{lemma}\label{xDEGamma}
The operators $X,D,E,\Delta$ and $\Gamma$
satisfy the following relations:
\begin{eqnarray*}
\begin{array}{lll}
\{ X,D \}=-2E-nI; & [E,D]=-D; & [E,X]=X  \\
\left[\Delta,X\right]=2D; &
XD=-E-\Gamma; & \left[D,X^2\right]=-2X
 \\
\left[E,X^2\right]=2X^2; & \left[E,\Delta \right]=-2\Delta; & [\Delta,X^2]=-4E-2nI\\
 \Gamma=E+nI+DX; & \{ \Gamma,X \}=(n-1)X; & \{ \Gamma,D \}=(n-1)D \\
 \left[\Gamma,X^2\right]=0; &  \left[\Gamma,\Delta\right]=0; &\left[E,\Gamma\right]=0. \\

\end{array}
\end{eqnarray*}
\end{lemma}
The proof of that statement can be found in \cite{DSS92}, Chapter II.

The relation $\left[ \Gamma,X^2\right]=0$ implies that $\Gamma f(|x|)=0$ while the relation $\left[\Gamma,\Delta\right]=0$ implies that $\Gamma (\ker \Delta) \subset \ker \Delta.$
Moreover, a full description
of $\mathcal{A}_n$ as a representation of the orthosymplectic Lie
algebra $\mathfrak{osp}(1|2)=\mathfrak{osp}(1|2)^{\small{\mbox{even}}}\bigoplus_{[\cdot,\cdot]} \mathfrak{osp}(1|2)^{\small{\mbox{odd}}}$ naturally follows by taking the generators $P^+,P^-,Q,R^+,R^-$ as follows:
\begin{eqnarray*}
\begin{array}{ccccc}
P^-=-\frac{1}{4}\Delta, & P^+=\frac{1}{2}X^2, & Q=\frac{1}{2}\left(E+\frac{n}{2}I\right), & R^+=iX,&R^-=i D. 
\end{array}
\end{eqnarray*}
Recall that the orthosymplectic Lie superalgebra $\mathfrak{osp}(1|2)$ (cf.\ \cite{FSS00}) has three even generators $P^+,P^-,Q \in \mathfrak{osp}(1|2)^{\small{\mbox{even}}}$ and two odd generators $R^+,R^- \in \mathfrak{osp}(1|2)^{\small{\mbox{odd}}}$ satisfying
the following commuting relations
\begin{eqnarray}
\label{LieSuperalgebraRelations}
\begin{array}{llll}
\left[ R^+, P^+\right]=0; & \left[ R^+, P^-\right]=\frac{1}{2}R^-; & [Q,R^+]=\frac{1}{2}R^+\\
\left[ R^-, P^+\right]=-R^+; & \left[ R^-, P^-\right]=0; & [Q,R^-]=-\frac{1}{2}R^- \\
\left[ P^-, P^+\right]=Q; & \left[ Q,
P^+\right]=P^+; & \left[ Q,
P^-\right]=-P^-.
\end{array}
\end{eqnarray}
Here we would like to stress that the even part of $\mathfrak{osp}(1|2)$ is isomorphic to the Lie algebra $\mathfrak{sl}_2(\BR)$, i.e. $\mathfrak{osp}(1|2)^{\mbox{\small{even}}}\cong \mathfrak{sl}_2(\BR)$.

\section{The Quantum Harmonic Oscillator}\label{HarmonicOscillator}

\subsection{Weyl-Heisenberg symmetries and Hermite expansions revisited}\label{WeylHeisenbergSymmetries}

In the sequel, we will consider the action of Hamiltonian operator $\mathcal{H}$ defined in (\ref{Hamiltonian}) on the configuration space $L_2(\BR^n)$. The usage of the dilation operator $S_{a}: f(x)\mapsto f(ax)$ allows us to re-scale the operator $\mathcal{H}$ in terms of the standard Hamiltonian $\mathcal{H}_0=\frac{1}{2} \left(-\Delta+|x|^2I\right)$ by getting rid of the constants $m$ and $\omega$ (cf.\ \cite{Folland08}, pages 53-56):
\begin{eqnarray}
\label{RescaleHamiltonian}S_{\sqrt{m \omega}}^{-1}\mathcal{H}S_{\sqrt{m \omega}}=\omega \mathcal{H}_0.
\end{eqnarray}
This shows that the eigenfunctions for the Hamiltonian operator (\ref{Hamiltonian}) can be described explicitly in terms of eigenfunctions of $\mathcal{H}_0$ using the Fock space formalism in phase space (\cite{Fock32};~\cite{Folland89}, pages 47--49).

We embed the Fock space over $L_2(\BR^n)$, say $\mathcal{F}(L_2(\BR^n))$, as a dense linear subspace generated from the Gaussian window $\phi(x)=\pi^{-\frac{n}{4}}e^{-\frac{|x|^2}{2}}$ and by the $2n+1$ elements $A_1^+,A_2^+,\ldots,A_n^+,A_1^-,A_2^-,\ldots,A_n^-,I$,
with
\begin{eqnarray}
\begin{array}{lll}
\label{WeylHeisenbergFock}A_j^+=\frac{1}{\sqrt{2}}\left(X_j-\partial_{X_j} \right), & A_j^-=\frac{1}{\sqrt{2}}\left( X_j+\partial_{X_j} \right).
\end{array}
\end{eqnarray}

From the relations (\ref{WeylHeisenberg}), it follows that the operators $A_j^\pm$ and $I$ generate the Weyl-Heisenberg Lie algebra $\mathfrak{h}_n$ (cf.\ \cite{Folland89}, Chapter 1).
On the other hand, straightforward computations show that the Gaussian window $\phi$ satisfies the following two properties on the configuration space $L_2(\BR^n)$. We have the
\begin{itemize}
\item Normalization property:

$\| \phi\|_{L_2(\BR^n)}^2=\langle \phi,\phi \rangle_{L_2(\BR^n)}=\pi^{-\frac{n}{2}}\int_{\BR^n}e^{-{|x|^2}} dx=1$.
\item Annihilation property:

    $A_j^-\phi(x)
    =\frac{1}{\sqrt{2}}\pi^{-\frac{n}{4}}
    \left(\frac{\partial}{\partial x_j}\left(e^{-\frac{|x|^2}{2}}\right)+x_je^{-\frac{|x|^2}{2}}  \right)=0.$
\end{itemize}

With the statements that we described above for $A_j^\pm$ and $\phi$, we can describe $\mathcal{F}(L_2(\BR^n))$ as a boson Fock space with $n$ particle states underlying the vacuum vector (the so-called ground state) $\phi$.
Moreover, a combination of the Weyl-Heisenberg character of $A_j^\pm$ and an application of mathematical induction on $d \in \BN$, results into the above referred commuting expression involving $A_j^-$ and the $d-$th power of $A_k^+$:
$$ \left[A_j^-,\left(A_k^+\right)^d\right]=d \left(A_k^+\right)^{d-1}.$$
Thus, according to the second quantization approach (cf.~\cite{Fock32}), the infinite set of vectors $\left\{\phi_\alpha\right\}_{\alpha \in (\BN_0)^n}$ labelled by the multi-index $\alpha=(\alpha_1,\alpha_2,\ldots,\alpha_n)$ such that
$$ \phi_\alpha(x)=\frac{1}{\sqrt{\alpha_1!\alpha_2!\ldots \alpha_n!}}\left(A_{1}^+\right)^{\alpha_1}\left(A_{2}^+\right)^{\alpha_2} \ldots \left(A_{n}^+\right)^{\alpha_n}\phi(x),$$ satisfy the following raising/lowering properties:
\begin{eqnarray}
\label{RaisingLowering}
\begin{array}{lll}
A_j^+ \phi_\alpha(x)=\sqrt{\alpha_j+1}~\phi_{\alpha+{\bf v}_j}(x), & A_j^- \phi_\alpha(x)=\sqrt{\alpha_j}~\phi_{\alpha-{\bf v}_j}(x),
\end{array}
\end{eqnarray}
respectively. Here ${\bf v}_j$ represents the $j-$canonical vector of $\BR^n$. This allows us to show that $\{ \phi_\alpha\}_\alpha$ is an orthonormal basis for $\mathcal{F}(L_2(\BR^n))$ (cf.\ \cite{Folland08}, page 54).
On the other hand, using the Lie algebra generators, we can also show that the standard Hamiltonian $\mathcal{H}_0=\frac{1}{2}(-\Delta+|x|^2I)$
and the number operator
\begin{eqnarray}
\label{numberOperator}E^{+-}=\sum_{j=1}^n A_j^+ A_j^-
\end{eqnarray} are interrelated by
$
\mathcal{H}_0=E^{+-}+\frac{n}{2}I.
$ On the basis of this property together with the relations (\ref{RaisingLowering}) we can show that $\left\{\phi_\alpha\right\}_{\alpha}$ is a set of eigenvectors for $\mathcal{H}_0$ corresponding to the eigenvalues $\epsilon_n\left(\alpha\right)=\sum_{j=1}^n \alpha_j+\frac{n}{2}$.  Property (\ref{RescaleHamiltonian}) implies that $S_{\sqrt{m \omega}}\phi_\alpha(x)=\phi_\alpha(\sqrt{m \omega}x)$ are solutions of the eigenvalue problem
$$ \mathcal{H} f(x)=\omega\epsilon_n(\alpha)~f(x).$$

The relation with the Weyl-Heisenberg Lie group $\mathbb{H}^n$ can be recast in terms of displacement operators by an exponentiation map  $\rho(t,{\bf \omega}):\mathfrak{h}_n \rightarrow \mathbb{H}^n $ as follows (cf.\ \cite{Folland89}, Chapter 1; \cite{Thangavelu93}, Section 1.2).
Recall that the Heisenberg group $\mathbb{H}^n$ is the non-commutative group represented on $\BR \times \BC^n$ in which the group multiplication is defined by $$(s,{\bf w}) * (t,{\bf z})=\left(s+t+i \Im ({\overline{{\bf w}} \cdot {{\bf z}}}),{\bf w}+{\bf z}\right).$$ The multiplicative inverse element then has the form $(s,{\bf w})^{-1}=(-s,-{\bf w})$.

For ${\bf w}=(w_1,w_2,\ldots,w_n) \in \BC^n$, with $w_j \in \BC$, we set ${\bf A}^\pm=(A_1^\pm,A_2^\pm,\ldots,A^\pm_n)$ and we define formally the inner product ${\bf w} \cdot {\bf A}^\pm$ as ${\bf w} \cdot {\bf A}^\pm=\sum_{j=1}^n w_j A_j^\pm$ and set $d \rho(t,{\bf w})=itI+ {\bf w} \cdot {\bf A}^+ - \overline{{\bf w}} \cdot {\bf A}^+ \in \mathfrak{h}_n$, $\rho(t,{\bf w})=\exp\left( d \rho(t,{\bf w}) \right)\in \mathbb{H}^n$.

It is easy to see that $\rho(t,{\bf w})=e^{it}\rho(0,{\bf w})$. On the other hand, the Baker-Campbell-Haussdorf formula (\cite{Perelomov86}, page 9) tells us that
\begin{eqnarray}
\label{BakerCambpellHaussdorf}\exp(U)\exp(V)=\exp\left(\frac{1}{2}[U,V]\right)\exp(U+V)
\end{eqnarray} whenever $\left[U,\left[U,V\right]\right]=0=\left[V,\left[U,V\right]\right]$. From this formula we may infer that the mapping
$${\bf w} \mapsto \rho(0,{\bf w})=\exp\left( {\bf w} \cdot {\bf A}^+ - \overline{{\bf w}} \cdot {\bf A}^-\right)$$
 is a projective representation on $\mathbb{H}^n$.
Indeed for any ${\bf w},{\bf z} \in \BC^n$, the relation $\left[{\bf w} \cdot {\bf A}^+- \overline{{\bf w}} \cdot {\bf A}^-,{\bf z} \cdot {\bf A}^+- \overline{{\bf z}} \cdot {\bf A}^-\right]=\left({\bf w}\cdot \overline{{\bf z}}-\overline{\bf w}\cdot {\bf z}\right)I$ leads to
\small{\begin{eqnarray*}
\rho(0,{\bf w})\rho(0,{\bf z})
=\exp\left(\frac{{\bf w}\overline{{\bf z}}-\overline{\bf w}{\bf z}}{2}~I\right)\rho(0,{\bf w}+{\bf z})=e^{-i \Im ({\overline{{\bf w}} \cdot {{\bf z}}})}\rho(0,{\bf w}+{\bf z}),
\end{eqnarray*}}
and hence, the homomorphism properties described below on $\mathbb{H}^n$ follow straightforwardly:
\begin{eqnarray*}
\rho((s,{\bf w})*(t,{\bf z}))=\rho(s,{\bf w})\rho(t,{\bf z}), & \mbox{and}~\rho((s,{\bf w})^{-1})=\rho(s,{\bf w})^{-1}.
\end{eqnarray*}

So, we can now introduce Perelomov coherent states (cf.\ \cite{Perelomov86,Wunsche91}) as the coherent states generated by the action of the operator $\rho(0,{\bf w})$ on the ground state $\phi(t)$, i.e. $\Phi_{{\bf w}}=\rho(0,{\bf w})\phi.$

By straightforward computations using the Baker-Campbell-Haussdorf formula (\ref{BakerCambpellHaussdorf}) and the identity $\exp(-\overline{{\bf w}}\cdot {\bf A}^-)\phi=\phi$, we can establish that
$$ \Phi_{{\bf w}} =e^{-\frac{|{\bf w}|^2}{2}} \exp({\bf w}\cdot {\bf A}^+)\exp(-\overline{{\bf w}}\cdot {\bf A}^-)\phi=e^{-\frac{|{\bf w}|^2}{2}}\sum_{\alpha=(\alpha_1,\alpha_2,\ldots,\alpha_n)\in \BN^{n}_0}\frac{w_1^{\alpha_1}w_2^{\alpha_2}\ldots w_n^{\alpha_n}}{\alpha_1!\alpha_2!\ldots \alpha_n!}  \phi_\alpha.$$

Here, we would like to remark that $\Phi_{\bf w}(t)$ is the so-called Segal-Bargmann kernel on $L_2(\BR^n)$ that allows us to express the Bargmann transform $\mathcal{B}=\exp\left(\frac{1}{2}|{\bf w}|^2\right)\left\langle \cdot, \Phi_{\overline{{\bf w}}} \right\rangle_{L_2(\BR^n)}$ in the theory of Segal-Bargmann spaces of analytic functions (cf.\ \cite{Bargmann61,Segal63}).

Summarizing, the flexibility of this approach based on the group-theoretical backdrop allows us to extend the construction of Perelomov coherent states to more general coherent states by means of the action of $\rho(0,{\bf w})$ on the states $\phi_\gamma$ (cf.\ \cite{Perelomov86}, Chapter 2). These operators are known in literature as displacement operators \cite{Wunsche91} or Heisenberg-Weyl operators \cite{DeGossonLuef09} underlying the so-called Weyl transform (cf.\ \cite{Thangavelu93}, Section 1.1).

 These produce coherent states $\Phi_{{{\bf w}},\gamma}=\rho(0,{\bf w})\phi_\gamma$ that allow us to express the so-called true poly-Bargmann transforms $$\mathcal{B}^\gamma=e^{\frac{|{\bf w}|^2}{2}}\left\langle \cdot, \Phi_{\overline{{\bf w}},\gamma} \right\rangle_{L_2(\BR^n)}$$ in the theory of Poly-Fock spaces or generalized Segal-Bargmann spaces. It also sheds some light on its interplay with the Gabor's windowed Fourier transform underlying Hermite windows. However, we shall not explore these relations here in depth. We refer for instance to \cite{AIM00,Vasilev00,AbreuACHA10,AbreuMon10,BrackenWatson10} and the references given therein to get a more complete overview of that topic.

\subsection{$\mathfrak{osp}(1|2)$ symmetries {\it vs.} Spectra of the Harmonic Oscillator}

In the sequel, we will consider the ladder operators $D^\pm$ that belong to $\mathcal{A}_n$ (see Subsection \ref{ospClifford}):
\begin{eqnarray}
\label{Dpm}
\begin{array}{llllll}
D^+&=&\frac{1}{\sqrt{2}}(X-D)&=&\sum_{j=1}^n\xi_j A^+_j \\
D^-&=&\frac{1}{\sqrt{2}}(X+D)&=&\sum_{j=1}^n \xi_j A^-_j,
\end{array}
\end{eqnarray}
where $A_j^\pm$ are defined by (\ref{WeylHeisenbergFock}).

From the definition, it follows that $D^+,D^-,\mathcal{H}_0\in \mbox{End}(\mathcal{S}(\BR^n;\BR_{0,n}))$ satisfy the anti-commuting relation
\begin{eqnarray}
\label{HamiltonianSplitting}\{ D^+,D^-\}=-2\mathcal{H}_0,
\end{eqnarray}
or equivalently, $\{ D^+,D^-\}=-2E^{+-}-nI$, in terms of the number operator $E^{+-}$ given by (\ref{numberOperator}).

In order to apply the Fock space formalism, we will take:
\begin{itemize}
\item The $\BR_{0,n}-$Hilbert module $L_2(\BR^n;\BR_{0,n})=L_2(\BR^n) \bigotimes \BR_{0,n}$ endowed with the bilinear form
$$ \langle f,g \rangle=\int_{\BR^n} f(x)^\dag g(x) dx,$$
where $dx$ stands for the Lebesgue measure over $\BR^n$.
\item The Clifford algebra-valued Schwartz spaces $\mathcal{S}(\BR^n;\BR_{0,n}):=\mathcal{S}(\BR^n)\bigotimes \BR_{0,n}$ on which $A_j^\pm$ and $D^\pm$ act as left endomorphisms.
\item The Clifford algebra-valued polynomial space $\mathcal{P}=\BR[\underline{x}]\bigotimes \BR_{0,n}$, where $\BR[\underline{x}]$ denotes ring of multivariate real-valued polynomials over $\BR^n$.
\end{itemize}

The following lemma from which one can deduce the Rodrigues's formula in special function theory provides us with the necessary motivation for the construction of eigenspaces for the Hamiltonian operator (\ref{Hamiltonian}) in terms of Clifford algebra-valued functions:
\begin{lemma}[See \ref{HarmonicOscillatorAppendix}]\label{RodriguesFormulae}
The operators $A_j^+,A_j^- \in \mbox{End}(\mathcal{S}(\BR^n;\BR_{0,n}))$ and $D^+,D^- \in \mbox{End}(\mathcal{S}(\BR^n;\BR_{0,n}))$ can be represented by
\begin{eqnarray}\label{AjexpX2}
\begin{array}{lllll}
A_j^+ &=& -\frac{1}{\sqrt{2}}\exp\left( -\frac{1}{2} X^2\right)\partial_{X_j}\exp\left(\frac{1}{2} X^2\right)\\ A_j^-&=&\frac{1}{\sqrt{2}}\exp\left(\frac{1}{2} X^2\right)\partial_{X_j}\exp\left(-\frac{1}{2} X^2\right)
\end{array} \\ \nonumber \\
\label{DexpX2}
\begin{array}{lllll}
D^+ &=& -\frac{1}{\sqrt{2}}\exp\left( -\frac{1}{2} X^2\right)D\exp\left(\frac{1}{2} X^2\right)\\ D^-&=&\frac{1}{\sqrt{2}}\exp\left(\frac{1}{2} X^2\right)D\exp\left(-\frac{1}{2} X^2\right).
\end{array}
\end{eqnarray}
\end{lemma}

The construction of Fock spaces on the Hilbert module $L_2(\BR^n;\BR_{0,n})$ can be performed in the same manner:

Firstly, we would like to stress that for each $f,g \in \mathcal{S}(\BR^n;\BR_{0,n})$ the order estimate $\left|f(x)^\dag g(x)\right|=O\left(|x|^{-2N}\right)$ for $N$ sufficiently large ensures that $\int_{\BR^n} D(f(x)^\dag g(x)) dx=0$. Hence, an application of the Clifford $\dag-$conjugation combined with the Leibniz rule over $\BR^n$ gives the following identity:
\begin{eqnarray}
\langle D^-f,g \rangle&=&\frac{1}{\sqrt{2}}\int_{\BR^n} ((X+D)f(x))^\dag g(x) dx\nonumber \\
&=&\frac{1}{\sqrt{2}}\int_{\BR^n} \left(-Xf(x)^\dag-Df(x)^\dag\right) g(x) dx\nonumber \\
\label{dualProperty}&=& -\frac{1}{\sqrt{2}}\int_{\BR^n} f(x)^\dag Xg(x) dx-\frac{1}{\sqrt{2}}\int_{\BR^n} Df(x)^\dag g(x) dx \\
&=&-\frac{1}{\sqrt{2}}\int_{\BR^n} f(x)^\dag g(x) dx+\frac{1}{\sqrt{2}}\int_{\BR^n} f(x)^\dag Dg(x) dx \nonumber \\
&=&\frac{1}{\sqrt{2}}\int_{\BR^n} f(x)^\dag(-X+D) g(x) dx\nonumber \\
&=&-\langle f,D^+g \rangle.\nonumber
\end{eqnarray}
This shows that, up to a minus sign, $D^+$ is the adjoint of $D^-$ in $L_2(\BR^n;\BR_{0,n})$. Hence, from (\ref{DexpX2}) Lemma \ref{RodriguesFormulae} $D^+$ is the adjoint of $\frac{1}{\sqrt{2}}D$ with respect to the $\BR_{0,n}-$valued bilinear form:
\begin{eqnarray}
\label{IntegralRepresentationFischer}
\langle f,g \rangle_{\mathcal{F}}=\left\langle \phi~ f,\phi~ g \right\rangle=\pi^{-\frac{n}{2}}\int_{\BR^n} f(x)^\dag g(x) e^{-|x|^2} dx.
\end{eqnarray}

Here we would like to stress that $\langle \cdot,\cdot \rangle_{\mathcal{F}}$ shall be regarded as the integral representation for the Fischer inner product (cf.\ \cite{DSS92}, pp. 204-205) that extends the integral representation obtained in \cite{Bargmann61,Segal63} by Bargmann and Segal. Moreover, the spaces $\mathcal{F}$ and $\mathcal{F}_k$ defined below:
\begin{eqnarray}
\label{FockSpace}\mathcal{F}=\left\{ f \in \ker D~:~\langle f,f \rangle_{\mathcal{F}}=\pi^{-\frac{n}{2}}\int_{\BR^n} |f(x)|^2 e^{-{|x|^2}} dx <\infty \right\} \\
\label{FockSpacek}\mathcal{F}_k=\left\{ f \in \ker D^k~:~\langle f,f \rangle_{\mathcal{F}}=\pi^{-\frac{n}{2}}\int_{\BR^n} |f(x)|^2 e^{-{|x|^2}} dx <\infty \right\}
\end{eqnarray}
shall be interpreted as the monogenic counterparts for the real Bargmann spaces (also called Segal-Bargmann, Fock or Fischer spaces, see \cite{Segal63,Bargmann61,NewmanShapiro66,CnopsKisil99}) and poly-Bargmann spaces (also called Poly-Fock or generalized Bargmann spaces, \cite{Vasilev00,AIM00}), respectively. This sort of spaces are proper subspaces of the so-called poly-monogenic functions with respect to the $C^\infty$-topology (cf.\ \cite{MR02}).
On the other hand, from Lemma \ref{RodriguesFormulae} and from the relation (\ref{DexpX2}), it follows that $D^-$ annihilates $\phi(x)f(x)$ for any $f \in ker D$:
\begin{eqnarray}
\label{DpmMonogenic}
\begin{array}{lll}
D^-\left(\phi(x)f(x)\right)&=&\pi^{-\frac{n}{4}}D^-\left(e^{-\frac{|x|^2}{2}}f(x)\right)\\
&=&\pi^{-\frac{n}{4}}\exp\left(\frac{1}{2} X^2\right)~Df(x)\\
&=&0.
\end{array}
\end{eqnarray}
Since the Gaussian window $\phi(x)=\pi^{-\frac{n}{4}}e^{-\frac{|x|^2}{2}}$ satisfies $\langle \phi,\phi \rangle=1$ (the normalization property), and the spectrum of $\mathcal{H}_0$ corresponds to the increasing sequence $\{k+\frac{n}{2}\}_{k \in \BN_0}$, the construction of the Fock states $\{ \Psi_k\}_{k \in \BN_0}$ viz $\Psi(x)=\phi(x)$ and $\Psi_k(x)=\frac{1}{\sqrt{c_k}}(D^+)^{k}\Psi(x)$, for some constant $c_k$, should take into account the constraints below:
\begin{itemize}
\item {\it Raising property:}  $D^+ \Psi_k(x)=\sqrt{k+1+\frac{n-1}{2}} \Psi_{k+1}(x)$ for each $k \in \BN$.
\item {\it Lowering property:} $D^- \Psi_0(x) =0$ and $D^- \Psi_k(x)=-\sqrt{k+\frac{n-1}{2}} \Psi_{k-1}(x)$ for each $k\in \BN$.
\end{itemize}

We will start by proving the following statements:

\begin{lemma}[See \ref{HarmonicOscillatorAppendix}]\label{powersDpm}
The operators $D^-,D^+ \in \mbox{End}(\mathcal{S}(\BR^n;\BR_{0,n}))$ satisfy
\begin{eqnarray}
\label{DminusDplusk}
\begin{array}{lll}
\left[D^-,(D^+)^k\right]=\left\{
\begin{array}{lll}
-2j (D^+)^{2j-1} &, \mbox{if}~k=2j \\ \ \\
2 (D^+)^{2j}\left( \Gamma-\left(\frac{n}{2}+j\right)I\right) &, \mbox{if}~k=2j+1
\end{array}
\right.
\end{array}
\end{eqnarray}
\end{lemma}

Next, for any starlike domain $\Omega$ with center $0$, we define for each $s>0$ the operator $I_s: C^1(\Omega;  \BR_{0,n})\longrightarrow
C^1(\Omega; \BR_{0,n})$ by
\begin{equation}\label{Is}
I_s f(\underline{x})=\int_0^1 f(t\underline{x}) t^{s-1} dt.
\end{equation}
Furthermore, we will write $I$ instead of $I_0$ to denote the identity operator.
The next lemma proved in \cite{MR02} shows that $I_s$ is the inverse of the operator $E_s:=E+sI$:

\begin{lemma}[\cite{MR02}, Lemma 3.1]\label{MalonekRenIs}
 Let $\underline{x}\in \BR^n$ and $\Omega$ be a domain with
$\Omega\supset [0, \underline{x}]$.  If $s>0$ and $f\in C^1(\Omega;
\BR_{0,n})$, then
\begin{equation}\label{eq1.880} f(\underline{x})=I_s E_{s} f (\underline{x})= E_{s} I_s f(\underline{x}).\end{equation}
\end{lemma}

Roughly speaking, from $\{X,D\}=-2E-nI=-2(E+\frac{n}{2}I)$ (see Lemma \ref{xDEGamma}), the mapping $f \mapsto -\frac{1}{2}I_{\frac{n}{2}}f$ shall be interpreted as a sort of right inverse for $DX$ on the range $\ker D$.
On the other hand, from $\left[E+\frac{n}{2}I,D\right]=-D$ (see relations (\ref{LieSuperalgebraRelations}) for the elements $R^+$ and $Q$), we obtain $D\left(E+\left(\frac{n}{2}+j\right)I\right)=\left(E+\left(\frac{n}{2}+j+1\right)I\right)D$ and hence
\begin{eqnarray}
\label{DIs}D I_{\frac{n}{2}+j}=I_{\frac{n}{2}+j+1}\left(E+\left(\frac{n}{2}+j+1\right)I\right)D I_{\frac{n}{2}+j}=I_{\frac{n}{2}+j+1}D.
\end{eqnarray}
This shows that the family of maps
$I_{\frac{n}{2}+j}: C^1(\Omega; \BR_{0,n})\longrightarrow
C^1(\Omega; \BR_{0,n})$ leave $\ker D$ invariant.

We have now the key ingredients to construct the eigenspaces for the Hamiltonian operator $\mathcal{H}_0$ in terms of Clifford algebra-valued functions:

For any $k \in \BN,$ let $U_0=I$ and $U_k=T_k T_{k-1} \ldots T_1$ with
\begin{eqnarray}
\label{TminusDplusk}T_k=\left\{
\begin{array}{lll}
(\frac{n-1}{4j}+1)I &, \mbox{if}~k=2j \\ \ \\
\left(\frac{n-1}{4}+j+\frac{1}{2}\right)~I_{\frac{n}{2}+j}&, \mbox{if}~k=2j+1
\end{array}
\right.
\end{eqnarray}

The theorem below shows that the Clifford-valued function spaces
$$\mathcal{F}_k^{+-}=\left\{ f_k(x)=\frac{1}{\sqrt{c_k}}(D^+)^k \left(\Psi_k(x)\right):\Psi_k(x)=\Psi(x)U_k(f(x)),f \in \ker D, ~ \langle \Psi_k,\Psi_k  \rangle=1  \right\},$$
with $c_k=\left(k+\frac{n-1}{2}\right)\left(k-1+\frac{n-1}{2}\right)\ldots\left(1+\frac{n-1}{2}\right)\frac{n-1}{2}$, contain the eigenfunctions of
$\mathcal{H}_0$ with eigenvalue $k+\frac{n}{2}$.

The following sequence of results will be useful for the characterization of $\mathcal{F}_k^{+-}:$

\begin{theorem}\label{RaisingLoweringFk}
For each $k \in \BN$, the functions $f_k \in \mathcal{F}^{+-}_{k}$ satisfy the following raising and lowering properties:
\begin{eqnarray*}
\begin{array}{ll}
D^+ f_k(x)=\sqrt{k+1+\frac{n-1}{2}}f_{k+1}(x), & D^- f_k(x)=-\sqrt{k+\frac{n-1}{2}}f_{k-1}(x).
\end{array}
\end{eqnarray*}

Moreover, they are solutions of the eigenvalue problem
$$ \mathcal{H}_0 f_k(x) =\left(k+\frac{n}{2}\right)f_k(x).$$
\end{theorem}

\proof
Let $f_k(x)=\frac{1}{\sqrt{c_k}}(D^+)^k \left(\Psi_k(x)\right)\in \mathcal{F}_k^{+-}$.
 From the definition of $\mathcal{F}^{+-}$, the raising property $D^+ f_k(x)=\sqrt{k+1+\frac{n-1}{2}}f_{k+1}(x)$ follows naturally:
\begin{eqnarray*}
D^+ f_k(x)&=&\frac{1}{\sqrt{c_k}}(D^+)^{k+1} \left(\Psi_k(x)\right)\\
&=&\sqrt{k+1+\frac{n-1}{2}}\frac{1}{\sqrt{\left(k+1+\frac{n-1}{2}\right)c_k}}(D^+)^{k+1} \left(\Psi_k(x)\right)\\
&=&\sqrt{k+1+\frac{n-1}{2}}~f_{k+1}(x).
\end{eqnarray*}

For the proof of the lowering property, we will combine Lemma \ref{powersDpm} with (\ref{DpmMonogenic}):

 Observe that the family of mappings $U_k$ defined by (\ref{TminusDplusk}) leave $\ker D$ invariant. Thus $U_k (f(x))$ is monogenic and hence, from (\ref{DpmMonogenic}) $\Psi_k(x)=\Psi(x) U_k (f(x))$ is a null solution of $D^-=\frac{1}{\sqrt{2}}(X+D)$:
$$D^{-}(\Psi_k(x))=\pi^{-\frac{n}{4}}D^{-}\left(e^{-\frac{|x|^2}{2}}  U_k (f(x))\right)=0.$$

Then, taking into account the recursive relation $U_{2j}=(\frac{n-1}{4j}+1)U_{2j-1}$. An application of Lemma \ref{powersDpm} then gives
\begin{eqnarray*}
D^-f_{2j}(x)&=&-\frac{2j}{\sqrt{c_{2j}}}(D^+)^{2j-1}(\Psi_{2j}(x))\\
&=&-\sqrt{2j+\frac{n-1}{2}}\sqrt{\frac{2j+\frac{n-1}{2}}{c_{2j}}}(D^+)^{2j-1}(\Psi(x)(U_{2j-1}f(x)))\\
&=&-\sqrt{2j+\frac{n-1}{2}}\frac{1}{\sqrt{c_{2j-1}}}(D^+)^{2j-1}(\Psi_{2j-1}(x)) \\
&=&-\sqrt{2j+\frac{n-1}{2}}~f_{2j-1}(x).
\end{eqnarray*}

 For $k=2j+1$ ($k$ odd), we show the lowering property by using the relation $\Gamma=-XD -E$ (see Lemma \ref{xDEGamma}), the recursive relation $U_{2j+1}=\left(\frac{n-1}{4}+j+\frac{1}{2}\right)~I_{\frac{n}{2}+j}U_{2j}$ and the property $\Gamma(\Psi(x)f(x))=\Psi(x)\Gamma f(x)$ (radial character of Gamma operator).

This results into
\begin{eqnarray*}
D^-f_{2j+1}(x)&=&
\frac{2}{\sqrt{c_{2j+1}}} (D^+)^{2j}\left( \Gamma-\left(\frac{n}{2}+j\right)I\right)\left(  \Psi_{2j+1}(x)\right)\\
&=& \frac{2j+1+\frac{n-1}{2}}{\sqrt{c_{2j+1}}} (D^+)^{2j}\left( \Gamma-\left(\frac{n}{2}+j\right)I\right)\left(  \Psi(x) I_{\frac{n}{2}+j} U_{2j} f(x)\right) \\
&=& \frac{2j+1+\frac{n-1}{2}}{\sqrt{c_{2j+1}}} (D^+)^{2j}\left(  \Psi(x) \left( \Gamma-\left(\frac{n}{2}+j\right)I\right)I_{\frac{n}{2}+j} U_{2j} f(x)\right) \\
&=& \sqrt{2j+1+\frac{n-1}{2}}\sqrt{\frac{2j+1+\frac{n-1}{2}}{c_{2j+1}}} \times \\
& &\times (D^+)^{2j}\left(  \Psi(x) \left( -E-\left(\frac{n}{2}+j\right)I\right)I_{\frac{n}{2}+j} U_{2j} f(x)\right) \\
&=& -\sqrt{2j+1+\frac{n-1}{2}}\frac{1}{\sqrt{c_{2j}}} (D^+)^{2j}\left(  \Psi(x) \left(  U_{2j} f(x)\right) \right) \\
&=&-\sqrt{2j+1+\frac{n-1}{2}}~f_{2j}(x).
\end{eqnarray*}

Thus, we have shown that $D^-f_{k}(x)=-\sqrt{k+\frac{n-1}{2}} f_{k-1}(x)$ holds for each $k \in \BN$.

Finally, from equation (\ref{HamiltonianSplitting}) we obtain that
\begin{eqnarray*}
\mathcal{H}_0 f_k =-\frac{1}{2}(D^-(D^+f_k)+D^+(D^-f_k))=\\
=\frac{1}{2}\left(\left(k+1+\frac{n-1}{2}\right)f_k+\left(k+\frac{n-1}{2}\right)f_k\right)=\left(k+\frac{n}{2}\right)f_k.
\end{eqnarray*}
\qed

We will conclude this subsection by showing that the mutual orthogonality between the spaces $\mathcal{F}_k^{+-}$ is obtained as a direct consequence of Theorem \ref{RaisingLoweringFk} and the adjoint property (\ref{dualProperty}). This provides us with a direct decomposition of the $\BR_{0,n}-$module $L_2(\BR^n;\BR_{0,n})$.

\begin{corollary}\label{orthonormalFk}
The spaces $\mathcal{F}_k^{+-}$ are mutually orthonormal with respect to the bilinear form $\langle \cdot,\cdot \rangle$, i.e.
$$ \mathcal{F}_k^{+-} \bot_{\langle \cdot,\cdot \rangle} \mathcal{F}_l^{+-}~~~\mbox{for}~k \neq l.$$
Each $\Psi_k \in \mathcal{F}^{+-}_k$ satisfies $\langle \Psi_k,\Psi_k \rangle=1$.

Moreover, the following direct decomposition of $L_2(\BR^n;\BR_{0,n})$ holds:
$$ L_2(\BR^n;\BR_{0,n})=\sum_{k=0}^\infty \oplus_{\langle \cdot,\cdot \rangle} \mathcal{F}_k^{+-}. $$
\end{corollary}

\proof
We first observe that $\langle \Psi,\Psi \rangle=1$ and $D^-(\Psi(x))=0$ follows from the construction. Then, the adjoint property (\ref{dualProperty}) implies that for each $k \in \BN$, $\Psi_k(x)=\frac{1}{\sqrt{c_k}}(D^+)^k(\Psi(x))$ is orthogonal to $\Psi(x)$ :
$$ \langle \Psi , \Psi_k \rangle=-\frac{1}{\sqrt{c_k}}\left\langle D^- \Psi, (D^+)^{k-1}(\Psi) \right\rangle=0.$$

Moreover, if $l \geq k>0$, then the combination of (\ref{dualProperty}) with the raising/lowering property for $f_k$ (Theorem \ref{RaisingLoweringFk}) leads to
\begin{eqnarray*}
\langle \Psi_l,\Psi_k \rangle&=&\frac{1}{\sqrt{(l+\frac{n-1}{2})(k+\frac{n-1}{2})}} \langle D^+\Psi_{l-1},D^+\Psi_{k-1}\rangle \\ &=&-\frac{1}{\sqrt{(l+\frac{n-1}{2})(k+\frac{n-1}{2})}} \langle D^-D^+\Psi_{l-1},\Psi_{k-1} \rangle \\
&=& \sqrt{\frac{l+\frac{n-1}{2}}{k+\frac{n-1}{2}}} \langle \Psi_{l-1},\Psi_{k-1} \rangle.
\end{eqnarray*}
By induction, the preceding calculation results into
\begin{eqnarray*}
\langle \Psi_l,\Psi_k \rangle=\sqrt{\frac{c_l}{c_k}} \langle \Psi,\Psi_{l-k} \rangle=\sqrt{\frac{c_l}{c_k}}\delta_{k,l}.
\end{eqnarray*}
We have proved the mutual orthonormality between the spaces $\mathcal{F}_k^{+-}$.

The statement of the direct decomposition of $L_2(\BR^n;\BR_{0,n})$ in terms of $\mathcal{F}_k^{+-}$ is then an immediate consequence following from the Fourier expansion for $f \in L_2(\BR^n;\BR_{0,n})$:
$$f =\sum_{k=0}^\infty \langle f,\Psi_k \rangle \Psi_k.$$
\qed

\subsection{Series Representation in terms of Clifford-Hermite functions or polynomials}\label{SeriesRepresentationCliffordHermite}

Before we proceed to the construction of series involving Clifford-Hermite functions and Clifford-Hermite polynomials, we will start to analyze the operator $\mathcal{H}_0=\frac{1}{2}\left( -\Delta+|x|^2I\right)$ by means of symmetries of $\mathfrak{sl}_2(\BR)$ (the even part of $\mathfrak{osp}(1|2)$).

We start with the following lemma, which relates the symmetries of $\mathcal{H}_0$ with the symmetries of the Hamiltonian $\mathcal{J}_0$:
\begin{eqnarray}
\label{HamiltonianJ0} \mathcal{J}_0=-\frac{1}{2}\Delta+E+\frac{n}{2}I,
\end{eqnarray}
and moreover, the symmetries of $\mathcal{J}_0$ with the symmetries of $E+\frac{n}{2}I$.

\begin{lemma}[see \ref{HarmonicOscillatorAppendix}]\label{H0Intertwining}
The operators $\mathcal{H}_0$, $\mathcal{J}_0$ and $E+\frac{n}{2}I$ are interrelated by
\begin{eqnarray}
\label{Statement1}
\begin{array}{lll}\exp\left(\frac{1}{2} X^2\right)\mathcal{J}_0&=&\mathcal{H}_0\exp\left(\frac{1}{2} X^2\right)\end{array}\\
\label{Statement2}
\begin{array}{lll}\exp\left(-\frac{1}{4}\Delta\right)\left( E+\frac{n}{2}I \right)&=&\mathcal{J}_0\exp\left(-\frac{1}{4}\Delta\right).\end{array}
\end{eqnarray}
\end{lemma}

Let us now restrict ourselves to the space of Clifford algebra-valued homogeneous polynomials of total degree $k$:
$$\mathcal{P}_k=\left\{ f \in \mathcal{P}~:~f(t\underline{x})=t^kf(\underline{x}),~\forall ~t \in \BR,\forall~ \underline{x}\in \BR^n\right\}.$$
Recall that we have a direct decomposition of $\mathcal{P}$ of the form:
\begin{eqnarray}
\mathcal{P}=\sum_{k=0}^\infty \bigoplus \mathcal{P}_k.
\end{eqnarray}
This follows from the fact that $\mathcal{P}_k$ are eigenspaces for the so-called Euler operator $E$ with eigenvalue $k$.
On the other hand, it is clear that $\mathcal{P}_k$ is a subset of the generalized poly-Bargmann space $\mathcal{F}_{k+1}$ (see (\ref{FockSpacek})). Notice that $D^{k+1}P_k(x)=0$ is true for each $k$ (cf.\ \cite{MR02}).

In the sequel, we also need to use the subspaces $\mathcal{P}_k \cap \ker \Delta$ (the so-called space of spherical harmonics of degree $k$) and $\mathcal{P}_k \cap \ker D$ (the so-called space of spherical monogenics of degree $k$). These spaces correspond to closed subspaces of $\mathcal{P}_k$ that satisfy one of the following coupled systems of equations, respectively:
\begin{eqnarray}
\label{sphericalHarmonics}\Delta f=0, & E f=kf. \\
\label{sphericalMonogenics}D f=0, & E f=kf.
\end{eqnarray}

Decompositions $\mathcal{P}_k$ in terms of spherical harmonics resp.\ monogenics of lower degrees are given by the Almansi/Fischer decomposition obtained in the book \cite{DSS92} and extended in \cite{MR02} for poly-harmonic resp.\ poly-monogenic functions of degree $k+1$ supported on star-like domains.

Recall that Fischer's decomposition (\cite{DSS92}, Theorem 1.10.1) gives a direct decomposition of $\mathcal{P}_k$ in terms of spherical monogenics of lower degrees:
$$ \mathcal{P}_k=\sum_{s=0}^{k} \bigoplus X^s \left(\mathcal{P}_{k-s} \cap \ker D \right)$$
while \cite{DSS92}, Corollary 1.3.3 gives a refinement of spherical harmonics in terms of spherical monogenics:
$$ \mathcal{P}_s \cap \ker \Delta =\left(\mathcal{P}_s \cap \ker D \right) \oplus X\left(\mathcal{P}_{s-1} \cap \ker D \right).$$
Moreover, each $P_k \in \mathcal{P}_k \cap \ker \Delta$ corresponds to $P_k(x)=M_k(x)+xM_{k-1}(x)$, where $M_k \in \mathcal{P}_k \cap \ker D$ and $M_{k-1} \in \mathcal{P}_{k-1} \cap \ker D$ are uniquely determined using the projection operators $\mathbb{P}$ and $\mathbb{Q}$:
\begin{eqnarray*}
\mathbb{P}=I+\frac{1}{2k+n-2}XD: \mathcal{P}_k \cap \ker \Delta \rightarrow \mathcal{P}_k \cap \ker D \\
\mathbb{Q}=-\frac{1}{2k+n-2}X D : \mathcal{P}_k \cap \ker \Delta \rightarrow X\left(\mathcal{P}_{k-1} \cap \ker D\right)
\end{eqnarray*}
i.e. $M_k(x)=\mathbb{P}(P_k(x))$ and $xM_{k-1}(x)=\mathbb{Q}(P_k(x))$.~Alternatively, in terms of the integral operators $I_{s}$ defined by (\ref{Is}) we have~$M_k(x)=(I+\frac{1}{2}XI_{\frac{n}{2}}D)P_k(x)$ and $xM_{k-1}(x)=-\frac{1}{2}XI_{\frac{n}{2}}(DP_k(x))$ (cf.\ \cite{MR02}).

The next theorem will give us the building blocks to construct Clifford-Hermite polynomials resp.\ functions in interplay with Sommen's approach from \cite{Sommen88}:
\begin{theorem}\label{eigenspacesH0J0}
For $k \in \BN_0$ and $\Psi(x)=\pi^{-\frac{n}{4}}e^{-\frac{|x|^2}{2}}$, define
\begin{eqnarray*}
\mathcal{P}_k^\Delta&=&\left\{ P_k^\Delta(x)=\exp\left(-\frac{1}{4}\Delta\right)P_k(x)~:~P_k \in \mathcal{P}_k\right\} \\
\mathcal{P}_k^{+-}&=&\left\{ P_k^{+-}(x)=\Psi(x)P_k^\Delta(x)~:~P_k^\Delta \in \mathcal{P}_k^\Delta\right\}.
\end{eqnarray*}
Then $\mathcal{P}_k^\Delta$ and $\mathcal{P}_k^{+-}$ are eigenspaces for $\mathcal{J}_0$ and $\mathcal{H}_0$ corresponding to the eigenvalue $k+\frac{n}{2}$.
\end{theorem}

\proof
Let $P_k \in \mathcal{P}_k$ and set $P_k^\Delta(x)=\exp\left( -\frac{1}{4}\Delta\right)P_k(x)$.

From the homogeneity of $P_k$, we obtain that $\left(E+\frac{n}{2}I\right)P_k(x)=\left(k+\frac{n}{2}\right)P_k(x)$. Next, a direct application of Lemma \ref{H0Intertwining} leads to
$$\mathcal{J}_0P_k^\Delta(x)=\exp\left( -\frac{1}{4}\Delta\right)\left(E+\frac{n}{2}I\right)P_k(x)=\left(k+\frac{n}{2}\right)P_k^\Delta(x)$$
$$\mathcal{H}_0P_k^{+-}(x)=\pi^{-\frac{n}{4}}\exp\left( \frac{1}{2} X^2\right)(\mathcal{J}_0P_k^\Delta(x))=\left(k+\frac{n}{2}\right)(\Psi(x)P_k^{\Delta}(x))=\left(k+\frac{n}{2}\right)P_k^{+-}(x).$$
This proves that $\mathcal{P}_k^\Delta$ and $\mathcal{P}_k^{+-}$ are eigenspaces for $\mathcal{J}_0$ and $\mathcal{H}_0$, respectively, with eigenvalue $k+\frac{n}{2}$.
\qed

Now we are able to establish a parallel between our approach and the approach obtained in \cite{Sommen88} by Sommen:

Firstly, we recall that the Cauchy-Kowaleskaya extension (see \cite{DSS92}, Subsection 5.1) of $f(x)$ in terms of the Cauchy-Riemann operator $\mathcal{D}=\frac{\partial}{\partial x_{n+1}}+\overline{\e_{n+1}}D$ corresponds to the solution of the Cauchy problem on $\BR \times [0,\infty)$ with initial data $F(x,0)=f(x)$:
\begin{eqnarray}\label{CKextension}
\left\{\begin{array}{ccc}
 \frac{\partial}{\partial x_{n+1}} F(x,x_{n+1})=-\overline{\e_{n+1}}DF(x,x_{n+1}) & \mbox{if}~x_{n+1}>0 \\ \ \\ F(x,0)=f(x), & \mbox{if}~x_{n+1}=0
\end{array}\right.
\end{eqnarray}
For $f(x)=P_k^\Delta(x)$, the Cauchy-Kowaleskaya extension results into the infinite series representation in terms of $M_k(x,x_{n+1})=\exp\left( -x_{n+1}\overline{\e_{n+1}} D\right)P_{k}(x) \in \ker \mathcal{D}$:
\begin{eqnarray*}
F(x,x_{n+1})=\exp\left( -x_{n+1}\overline{\e_{n+1}} D\right)P_{k}^\Delta(x)=\sum_{j=0}^\infty \frac{(-1)^j}{4^j j!}\Delta^j(M_k(x,x_{n+1}))
\end{eqnarray*}
The latter representation shall be understood as the series representation for the inversion of the Segal-Bargmann transform applied on the spaces of monogenic functions on $\BR^n \times [0,\infty)$ (cf.\ \cite{CnopsKisil99}).

 On the other hand, from $I_s(P_k(x))=\frac{1}{k+s}P_k(x)$, it is clear from the construction of $U_k$ by (\ref{TminusDplusk}) that $U_k(P_k(x))=u_k P_k(x)$ for some $u_k\in \BR$.

 So, on the basis of the integral representation (\ref{IntegralRepresentationFischer}) we can say that, if we take $P_k \in \mathcal{P}_k$ (an element of the space $\mathcal{F}_{k+1}$) such that $\langle P_k,P_k \rangle_{\mathcal{F}}=\frac{1}{u_k^2}$, then it is clear that $\Psi_k(x)=\Psi(x)U_{k}(P_k(x))$ is a normalized vector on $L_2(\BR^n;\BR_{0,n})$:
 $$\langle \Psi_k,\Psi_k\rangle=\langle U_{k}P_k,U_{k}P_k \rangle_{\mathcal{F}}=u_k^2\langle P_k,P_k\rangle_{\mathcal{F}}=1.$$

Moreover, from Theorems \ref{RaisingLoweringFk},~\ref{eigenspacesH0J0} and Corollary \ref{orthonormalFk}, the characterizations that we are going to recall now for Clifford-Hermite functions resp.\ polynomials are in fact rather obvious (cf.\ \cite{Sommen88}, Section 5):
\begin{itemize}
\item {\bf Normalized Clifford-Hermite functions of degree $k$:} The functions of the type $\Psi_k(x)=\frac{1}{\sqrt{c_k}}(D^+)^k(\Psi(x))$ that belong to $\mathcal{F}_k^{+-}$.
\item {\bf Normalized Clifford-Hermite polynomials of degree $k$:} The functions of the type $\psi_k(x)=\pi^{\frac{n}{4}}e^{\frac{|x|^2}{2}}\Psi_k(x)$ are eigenfunctions of the Hamiltonian $\mathcal{J}_0$ with eigenvalue $k+\frac{n}{2}$ are mutually orthonormal in the Fock space $\mathcal{F}$ (see (\ref{FockSpace})).
\item {\bf Generating functions:} The Cauchy-Kowaleskaya extension $$F(x,x_{n+1})=\exp(-x_{n+1}\overline{\e_{n+1}}D)\psi_k(x)$$ obtained from (\ref{CKextension}) gives a generating function in terms of Clifford-Hermite polynomials while $\Psi(x)F(x,x_{n+1})=\Psi(x)\exp(-x_{n+1}\overline{\e_{n+1}}D)\psi_k(x)$ gives a generating function in terms of Clifford-Hermite functions.
\end{itemize}

In conclusion, the Clifford-Hermite functions resp.\ polynomials in the above constructions are obtained in a combinatorial way by means of the Fock space formalism (cf.\ \cite{Fock32}) and, contrary to \cite{Sommen88}, this approach does not require {\it a priori} any knowledge of Cauchy's integral formula to ensure the mutual orthonormality of the Clifford-Hermite polynomials resp.\ functions.

\section{Solutions of the time-harmonic Maxwell equations with additional angular part} \label{MaxwellEquations}

\subsection{Symmetries and Series Representation of Solutions}\label{symmetriesMaxwell}

The main purpose of this section is to introduce some generalized
Clifford algebra valued operators that allow us to describe the
solutions of Maxwell's equations by means of the theory of
spherical monogenic resp.\ harmonic functions.

We start by proving the existence of an
isomorphism between the algebras of Clifford operators
$\mathcal{A}_n$ and
$$ \mathcal{A}^n_\lambda =\mbox{span}\left\{ X-\lambda I+\frac{2 \lambda}{n} \Gamma,D-\lambda I+\frac{2 \lambda}{n} \Gamma, \xi_j~:~j=1,\ldots,n \right\} $$
by means of the $\exp\left(\frac{\lambda}{n}(D-X)\right)$-action:

\begin{lemma}[see \ref{MaxwellEquationsAppendix}]\label{DisplacedIntertwine}
The action of the operator $\exp\left(\frac{\lambda
}{n}\left(D-X\right)\right)$ on $\mathcal{S}(\BR^n;\BR_{0,n})$ gives rise to
\begin{eqnarray*}
\exp\left(\frac{\lambda
}{n}\left(D-X\right)\right)~D=\left(D+\frac{2\lambda}{n}\Gamma-\lambda
I\right)\exp\left(\frac{\lambda
}{n}\left(D-X\right)\right) \\
\exp\left(\frac{\lambda
}{n}\left(D-X\right)\right)~X=\left(X+\frac{2\lambda}{n}\Gamma-\lambda
I\right)\exp\left(\frac{\lambda
}{n}\left(D-X\right)\right).
\end{eqnarray*}
\end{lemma}

The above established lemmas show that the operators
$D+\frac{2\lambda}{n}\Gamma-\lambda I$ and
$X+\frac{2\lambda}{n}\Gamma-\lambda I$ play the same role as the
standard classical operators $X$ and $D$, respectively. This is
due to the fact that the action of
$\exp\left(\frac{\lambda}{n}(D-X)\right)$ preserves the
(anti-)commutation relations.

Next, we will describe the series representation of the solutions
for PDEs of the type
\begin{eqnarray}
\label{PDEMaxwell}\left( D-\lambda+\frac{2\lambda}{n}\Gamma\right)f_\lambda=g_\lambda, & \mbox{with}~g_\lambda \in \ker \left(D-\lambda I+\frac{2\lambda}{n}\Gamma\right)^s.
\end{eqnarray}

To do so, we will apply the following multiplication rule on
$\mathcal{S}(\BR^n;\BR_{0,n}):$
$$\exp\left(\frac{\lambda}{n}(D-X)\right)=\exp\left(-\frac{1}{2} X^2\right)\exp\left(\frac{\lambda}{n}D\right)\exp\left(\frac{1}{2} X^2\right).$$
This equality follows from the relations (\ref{DexpX2}) and from
the series expansion of $\exp\left(\frac{\lambda}{n}D\right)$ in
the $C^\infty$-topology.

The next lemma provides us with the key ingredient to
compute the expressions $f_\lambda:=\exp(\frac{\lambda}{n}(D-X))f$
from $\mathcal{S}(\BR^n,\BR_{0,n})$:
\begin{lemma}[see \ref{MaxwellEquationsAppendix}]\label{expLambaDminusX}
When acting on $\mathcal{S}(\BR^n;\BR_{0,n})$, the multiplication rule
$$\exp\left(\frac{\lambda}{n}(D-X)\right)=\exp\left(-\frac{1}{2} X^2\right)\exp\left(\frac{\lambda}{n}D\right)\exp\left(\frac{1}{2} X^2\right)$$
is represented by the following series expansion that converges in the $C^\infty$-topology:
\begin{eqnarray*}
\exp\left(\frac{\lambda}{n}(D-X)\right)=\sum_{k=0}^\infty \frac{\lambda^{2k}}{n^{2k}(2k)!}\left(I+\frac{\lambda}{n(2k+1)}(D-X)\right)\left(2\mathcal{J}_0\right)^k.
\end{eqnarray*}

Here, $\mathcal{J}_0$ is the Hamiltonian operator defined by the equation (\ref{HamiltonianJ0}).
\end{lemma}

\begin{remark}
Notice that $\ker\left(D-\lambda I+\frac{2\lambda}{n}\Gamma\right)^0$ coincides with the trivial subspace $\{ 0\}$. Hence, using the series representation in the $C^\infty-$topology for $f_\lambda$ that is given in Lemma \ref{expLambaDminusX}, we can establish that
$$f_\lambda=\sum_{k=0}^\infty \frac{\lambda^{2k}}{n^{2k}(2k)!}\left(\left(2\mathcal{J}_0\right)^kf+\frac{\lambda}{n(2k+1)}(D-X)\left(2\mathcal{J}_0\right)^kf\right)$$ is a solution of (\ref{PDEMaxwell}) that belongs to $\ker \left( D-\lambda+\frac{2\lambda}{n}\Gamma\right)$ in the case when $s=0$.

Moreover, from Lemma \ref{DisplacedIntertwine} we can conclude that the above described series representation is a solution of (\ref{PDEMaxwell}) whenever $f$ belongs to $\mathcal{F}_{s+1}$ (see (\ref{FockSpacek})).
\end{remark}

The next theorem provides us with a meaningful characterization for the series representation for the solutions
of the PDE system (\ref{PDEMaxwell}):

\begin{theorem}\label{SeriesExpansionSolutions}
For each $P_s \in \mathcal{P}_s$, let $P_s^\Delta(x)=\exp\left(-\frac{1}{4}\Delta\right)P_s(x)$ and  $P_{\lambda,s}(x)=\exp\left(
\frac{\lambda}{n}(D-X)\right)P_s^\Delta(x)$.

Then $P_{\lambda,s}$ is a solution of the PDE system
(\ref{PDEMaxwell}) with the following properties:
\par\medskip\par
\noindent 1. $P_{\lambda,s}$ is a solution of the eigenvalue
problem
$$ \Delta (P_{\lambda,s}(x))-\frac{2\lambda}{n}X (P_{\lambda,s}(x)) =2E(P_{\lambda,s}(x))-2sP_{\lambda,s}(x).$$2. $P_{\lambda,s}$ is explicitly given by
\begin{eqnarray*}
P_{\lambda,s}(x)=\\
=\cosh\left(
\frac{\lambda}{n}\sqrt{2s+n}\right)P_s^{\Delta}(x)+\frac{1}{\sqrt{2s+n}}\sinh\left(
\frac{\lambda}{n}\sqrt{2s+n}\right)(D-X)P_s^{\Delta}(x)\nonumber \\
=\exp\left( -\frac{1}{4}\Delta\right) \times\\
\times \left(\cosh\left( \frac{\lambda}{n}\sqrt{2s+n}\right)P_s(x)+\frac{1}{\sqrt{2s+n}}\sinh\left( \frac{\lambda}{n}\sqrt{2s+n}\right)\left(\frac{D}{2}-X\right)P_s(x)\right).
\end{eqnarray*}
3. If $P_s \in \ker D$, then
\begin{eqnarray*}
P_{\lambda,s}(x)=\cosh\left( \sqrt{2s+n}\frac{\lambda}{n}\right)P_s(x)-\frac{1}{\sqrt{2s+n}}\sinh\left( \sqrt{2s+n}\frac{\lambda}{n}\right)x~P_s(x).
\end{eqnarray*}
is a solution of the coupled system of equations:
\begin{eqnarray*}
\left\{\begin{array}{ccc}
(D-\lambda I)f(x)=-\frac{2\lambda}{n}~\Gamma \left(f(x)\right) \\ \ \\ 2E(f(x))-2sf(x)=-\frac{2\lambda}{n}X f(x)
\end{array}\right.
\end{eqnarray*}
Moreover, $P_{\lambda,s}(x)$ is harmonic, i.e.
$P_{\lambda,s}\in\ker \Delta$.
\end{theorem}

\proof
We will prove the statements 1., 2. and 3. separately:
\\ \ \\
{\bf Proof of Statement 1:}
\\
Notice that for $P_s\in \mathcal{P}_s$, the expression $P^{\Delta}_s(x):=\exp\left(-\frac{1}{4}\Delta\right)P_s(x)$ satisfies
$$\mathcal{J}_0\left(P^{\Delta}_s(x)\right)=\left(s+\frac{n}{2}\right)P^{\Delta}_s(x),$$
where $\mathcal{J}_0$ is the Hamiltonian operator defined in (\ref{HamiltonianJ0}).

First we show that $\exp\left( \frac{\lambda}{n}\left( D-X\right)\right)\mathcal{J}_0=\left( \mathcal{J}_0+\frac{\lambda}{n}X \right)\exp\left( \frac{\lambda}{n}\left( D-X\right)\right)$. From Lemma \ref{xDEGamma}, we may infer that
\begin{eqnarray*}
\begin{array}{lll}
\left[ -\frac{1}{2}\Delta, X\right]=-D, &
\left[ E+\frac{n}{2}I,D\right]=-D, \\
\left[ E+\frac{n}{2}I,X\right]=X, &
\left[ D,-\frac{1}{2}\Delta\right]=0.
\end{array}
\end{eqnarray*}
This in turn leads to
\begin{eqnarray*}
\left[ \mathcal{J}_0,\frac{\lambda}{n}(D-X)\right]
=\\
=\frac{\lambda}{n}\left(\left[ -\frac{1}{2}\Delta,D\right]-\left[ -\frac{1}{2}\Delta,X\right]
+\left[ E+\frac{n}{2}I,D\right]-\left[ E+\frac{n}{2}I,X\right] \right) \\
\\
=\frac{\lambda}{n}\left( 0+D-D-X\right)\\
=-\frac{\lambda}{n}X,
\end{eqnarray*}
and hence,
$\left[ \mathcal{J}_0,
\exp\left(\frac{\lambda}{n}(D-X)\right)\right]=-\frac{\lambda}{n}X\exp\left(\frac{\lambda}{n}(D-X)\right).$

The latter equation is equivalent to
$$ \left(\mathcal{J}_0+\frac{\lambda}{n}X\right)\exp\left( \frac{\lambda}{n}\left( D-X\right)\right)=\exp\left( \frac{\lambda}{n}\left( D-X\right)\right)\mathcal{J}_0.$$
Therefore, for $P_{\lambda,s}(x)=\exp\left( \frac{\lambda}{n}\left(D-X\right)\right)P_{s}^{\Delta}(x)$ we obtain
$$ \left(\mathcal{J}_0+\frac{\lambda}{n}X\right)P_{\lambda,s}(x)
=\exp\left( \frac{\lambda}{n}\left( D-X\right)\right)\left(\mathcal{J}_0 P_{s}^{\Delta}(x)\right)
=\left(s+\frac{n}{2}\right)P_{\lambda,s}(x).$$
This is equivalent to
$$ \Delta (P_{\lambda,s}(x))-\frac{2\lambda}{n}X (P_{\lambda,s}(x)) =2E(P_{\lambda,s}(x))-2sP_{\lambda,s}(x).$$
\\
{\bf Proof of Statement 2:}
\\ From Lemma \ref{expLambaDminusX} we know that we can express $P_{\lambda,s}(x)=\exp\left( \frac{\lambda}{n}(D-X)\right)P_s^{\Delta}(x)$ as follows
\begin{eqnarray*}
P_{\lambda,s}(x)&=&\sum_{k=0}^\infty \frac{\lambda^{2k}}{n^{2k}k!}\left(\left(2\mathcal{J}_0\right)^k P_s^{\Delta}(x)+\frac{\lambda}{n(2k+1)}(D-X)\left(2\mathcal{J}_0\right)^k P_s^{\Delta}(x)\right)
\\
&=&\sum_{k=0}^\infty\frac{\lambda^{2k}(2s+n)^k}{n^{2k}k!}\left(P_s^{\Delta}(x)+\frac{\lambda}{n(2k+1)}(D-X) P_s^{\Delta}(x)\right).
\end{eqnarray*}
Next we apply the series expansion of the hyperbolic functions $t
\mapsto \cosh(t)$ and $t \mapsto \sinh(t)$. This allows us to
conclude that the latter expression is equivalent to the following
expression in the $C^\infty-$topology:
\begin{eqnarray}
\label{PLambdaS2b}P_{\lambda,s}(x)=\nonumber\\
=\cosh\left( \frac{\lambda}{n}\sqrt{2s+n}\right)P_s^{\Delta}(x)+\frac{1}{\sqrt{2s+n}}\sinh\left( \frac{\lambda}{n}\sqrt{2s+n}\right)(D-X)P_s^{\Delta}(x).
\end{eqnarray}
Using the relation $[\Delta,X]=2D$ (see Lemma \ref{xDEGamma}) and
applying induction arguments together with the series expansion
for $\exp\left( -\frac{1}{4}\Delta\right)$, allows us to establish
the relation
$$\left[X,\exp\left(-\frac{1}{4}\Delta\right)\right]=\frac{D}{2}
\exp\left(-\frac{1}{4}\Delta\right).$$ This is equivalent to
$\left(\frac{D}{2}-X\right)\exp\left(-\frac{1}{4}\Delta\right)=\exp\left(-\frac{1}{4}\Delta\right)(-X)$ and hence
$$\left(D-X\right)\exp\left(-\frac{1}{4}\Delta\right)
=\frac{D}{2}\exp\left(-\frac{1}{4}\Delta\right)+\left(\frac{D}{2}-X\right)\exp\left(-\frac{1}{4}\Delta\right)
=\exp\left(-\frac{1}{4}\Delta\right)\left(\frac{D}{2}-X\right).
$$

Therefore, equation (\ref{PLambdaS2b}) is equivalent to
\begin{eqnarray*}
P_{\lambda,s}(x)=\exp\left( -\frac{1}{4}\Delta\right) \times\\
\times \left(\cosh\left( \frac{\lambda}{n}\sqrt{2s+n}\right)P_s(x)+\frac{1}{\sqrt{2s+n}}\sinh\left( \frac{\lambda}{n}\sqrt{2s+n}\right)\left(\frac{D}{2}-X\right)P_s(x)\right).
\end{eqnarray*}
This completes the proof of statement 2.
\\
{\bf Proof of Statement 3:}
\\
Since $P_s\in \ker D$, we obtain $-\Delta P_s(x)=D(DP_s(x))=0$.
Therefore,
$$P^{\Delta}_s(x)=\exp\left(-\frac{1}{4}\Delta\right)P_s(x)=P_s(x).$$

Hence, the validity of the expression for $P_{\lambda,s}(x)$ in the following form
\begin{eqnarray*}
P_{\lambda,s}(x)=\cosh\left( \sqrt{2s+n}\frac{\lambda}{n}\right)P_s(x)-\frac{1}{\sqrt{2s+n}}\sinh\left( \sqrt{2s+n}\frac{\lambda}{n}\right)x~P_s(x)
\end{eqnarray*}
follows by inserting $P^{\Delta}_s(x)=P_s(x)$ into the equation
that we previously obtained in Statement 2.

Finally, the validity of the equation
$$2E(P_{\lambda,s}(x))-2sP_{\lambda,s}(x)=-\frac{2\lambda}{n}X
(P_{\lambda,s}(x))$$ follows by inserting $P^{\Delta}_s(x)=P_s(x)$
into the equation obtained in statement 1. The validity of the
equation $(D-\lambda I)P_{\lambda,s}(x)=-\frac{2\lambda}{n}\Gamma
P_{\lambda,s}(x)$ follows from Lemma \ref{DisplacedIntertwine}.

Moreover, the validity of the property $P_{\lambda,s}\in \ker
\Delta$ is a consequence from the property $[\Delta,X]=2D$ and
from the inclusion property $\ker D \subset \ker \Delta$. \qed

\subsection{Relation with Landau Operators}

In this subsection we will obtain a derivation of Landau operators
that describe symmetries in electromagnetism (cf.\ \cite{GHJ02})
in terms of $\mathfrak{osp}(1|2)$ symmetries. Moreover, we are
able to derive series representations that involve the solutions
of the time-harmonic Maxwell equations with angular part i.e. the
null solutions of $D-\lambda I+\frac{2\lambda}{n}\Gamma$.

In what follows we will exhibit the relations between
the spectrum of $\mathcal{H}_0$ and the Landau operator
$\mathcal{H}_\lambda=\tilde{\mathcal{H}}_\lambda+\mathcal{L}_\lambda$ .
Without loss of generality, let us consider for unit mass and frequency, the Landau operator (\ref{LandauOperator})
written in terms of the operators $D \mp \lambda I$, $X$ and $\Gamma$:
\begin{eqnarray}
\label{magneticLaplacian}\mathcal{H}_{\lambda}=\frac{1}{2}\left((D-\lambda I)(D+\lambda I)-X^2-\frac{2\lambda}{n}\left(X-2\lambda\left(I-\frac{1}{n}\Gamma\right)\Gamma\right) \right).
\end{eqnarray}
We start by deriving the analogues of $D^\pm$ in terms of the $\exp\left( \frac{\lambda}{n}D \right)-$action:

\begin{lemma}[see \ref{MaxwellEquationsAppendix}]\label{DpmIntertwining}
When acting on $\mbox{End}(\mathcal{S}(\BR^n;\BR_{0,n}))$, we obtain
\begin{eqnarray}
\label{DpmLambda}
\begin{array}{lll}
\exp\left(\frac{\lambda}{n}D\right) D^+\exp\left(-\frac{\lambda}{n}D\right)&=&\frac{1}{\sqrt{2}}\left(X- (D+{\lambda}I)+\frac{2 \lambda}{n}\Gamma\right) \\ \exp\left(\frac{\lambda}{n}D\right) D^- \exp\left(-\frac{\lambda}{n}D\right)&=& \frac{1}{\sqrt{2}}\left(X+ (D-{\lambda}I)+\frac{2 \lambda}{n}\Gamma\right).
\end{array}
\end{eqnarray}
\end{lemma}

Here we would like to stress that the isomorphism
$\exp\left(\frac{\lambda}{n}D \right)$ preserves the
(anti-)commuting relations, and hence, the
relations~(\ref{LieSuperalgebraRelations}) can be lifted from
$X,D$ and $E$ for the operators $X-\lambda I+
\frac{2\lambda}{n}\Gamma$, $D$ and
$E+\frac{n}{2}I+\frac{\lambda}{n}D$, respectively. This
corresponds to the proposition given below:
\begin{proposition}\label{ospMagnetic}
The elements $P^+,P^-,Q,R^+,R^-$ defined by
\begin{eqnarray*}
\begin{array}{lllll}
P^-=-\frac{1}{4}\Delta; & P^+=\frac{1}{2}\left(X-\lambda I+\frac{2\lambda}{n}\Gamma\right)^2; &
Q=\frac{1}{2}\left(E+\frac{n}{2}I\right)+\frac{\lambda}{2n}D \\
R^-=i D; & R^+=i\left(X-\lambda I+\frac{2\lambda}{n}\Gamma\right)&
\end{array}
\end{eqnarray*}
are the generators of the orthosymplectic Lie algebra $\mathfrak{osp}(1|2)$.
\end{proposition}

The next lemma shows that $\mathcal{H}_\lambda$ can be decomposed in terms
of the ladder operators $\frac{1}{\sqrt{2}}\left(X- (D+{\lambda}I)+\frac{2 \lambda}{n}\Gamma\right)$ and $\frac{1}{\sqrt{2}}\left(X+ (D-{\lambda}I)+\frac{2 \lambda}{n}\Gamma\right)$:

\begin{lemma}[see \ref{MaxwellEquationsAppendix}]\label{magneticLaplacianSplitting}
When acting on $\mbox{End}(\mathcal{S}(\BR^n;\BR_{0,n}))$, we have
$$ \left\{ \frac{1}{\sqrt{2}}\left(X- (D+{\lambda}I)+\frac{2 \lambda}{n}\Gamma\right) ,\frac{1}{\sqrt{2}}\left(X+ (D-{\lambda}I)+\frac{2 \lambda}{n}\Gamma\right) \right\}=-2\mathcal{H}_\lambda$$
\end{lemma}

\begin{remark}
In the spirit of \cite{ZKI10}, the operators
$\frac{1}{\sqrt{2}}\left(X- (D+{\lambda}I)+\frac{2
\lambda}{n}\Gamma\right)$ and \\ $\frac{1}{\sqrt{2}}\left(X+
(D-{\lambda}I)+\frac{2 \lambda}{n}\Gamma\right)$ may be
interpreted as hypercomplex extensions of the canonical creation
resp.\ annihilation operators on the quaternionic field which are
equivalent to $D^-$ resp.\ $D^+$ under the
$\exp\left(\frac{\lambda}{n}D\right)-$action.

In the spirit of \cite{DeGossonLuef09},
$\exp\left(\frac{\lambda}{n}D\right)$ plays the same role as the
wave-packet transform encoded in the cross-Wigner distribution. On
the other hand, from Proposition \ref{ospMagnetic}, the magnetic
Laplacian $\mathcal{H}_\lambda$ written in (\ref{magneticLapl})
can be obtained by the covariant action $X \mapsto
X-\lambda+\frac{2\lambda}{n}\Gamma$ on the spherical potential
$-\frac{1}{2}X^2=\frac{1}{2}|x|^2I$.
\end{remark}

Let us now focus our attention on the construction of
Clifford-Hermite functions resp.\ polynomials that we obtained in
Subsection \ref{SeriesRepresentationCliffordHermite} and on the
series representation for the solutions of the PDE system
(\ref{PDEMaxwell}) obtained in Subsection
\ref{symmetriesMaxwell}:\\ From Theorem
\ref{SeriesExpansionSolutions}, we can draw the conclusion that
$f_\lambda$ may be written as a generating function in terms of
Clifford-Hermite polynomials $\psi_k$:
\begin{eqnarray*}
\label{generatingCliffordHermite}f_\lambda(x)=\sum_{k=0}^\infty \frac{\lambda^k}{n^k~k!} \psi_k(x).
\end{eqnarray*}
This can be done following the same lines of the proof of
statement 2 from Theorem \ref{SeriesExpansionSolutions}. The
latter result shall be interpreted as the Bargmann inversion
formula in the space of Clifford algebra-valued polynomials
$\mathcal{P}$ (cf.\ \cite{CnopsKisil99}).

Moreover, since for each $\psi_k(x)$, the expression
$\mathbb{P}(\psi_k(x))=\psi_k(x)+\frac{1}{2k+n-2}xD(\psi_k(x))$ is
monogenic (cf.\ \cite{DSS92}, Corollary 1.3.3); each $f_\lambda
\in \ker\left(D-\lambda+\frac{2\lambda}{n}\Gamma\right)$ may be
represented in terms of the series expansion
$$f_\lambda(x)=\sum_{k=0}^\infty \frac{\lambda^k}{n^k~k!}  \left( \psi_k(x)+\frac{1}{2k+n-2}xD(\psi_k(x)) \right)$$
that corresponds to a generating function in terms of (monogenic) Clifford-Hermite polynomials.

Thus, by applying the multiplication
formula for $\exp\left( \frac{\lambda}{n}(D-X)\right)$:
\begin{center} $\exp\left( \frac{\lambda}{n}(D-X)\right)=
\exp\left(-\frac{1}{2} X^2\right)\exp\left( \frac{\lambda}{n}D\right)\exp\left(\frac{1}{2} X^2\right)$
\end{center}
and the characterization of Clifford Hermite functions resp.\ polynomials presented in Lemma \ref{eigenspacesH0J0}, we can deduce the next theorem.
\begin{theorem}
The following statements are true:
\begin{enumerate}
\item For each $\Psi_k \in \mathcal{F}_k^{+-}$ (see also Theorem \ref{RaisingLoweringFk}), the series expansion $$\Psi_{\lambda,k}(x):=\exp\left( \frac{\lambda}{n}D\right)\Psi_k(x)=\sum_{k=0}^\infty \frac{\lambda^k}{n^k~k!}D^k(\Psi_k(x))$$ satisfies the raising resp.\ lowering properties:
    \begin{eqnarray*}
    \frac{1}{\sqrt{2}}\left(X- (D+{\lambda}I)+\frac{2 \lambda}{n}\Gamma\right)\Psi_{\lambda,k}(x)&=&\sqrt{k+1+\frac{n-1}{2}}\Psi_{\lambda,k+1}(x) \\ \frac{1}{\sqrt{2}}\left(X+ (D-{\lambda}I)+\frac{2 \lambda}{n}\Gamma\right)\Psi_{\lambda,k}(x)&=&-\sqrt{k+\frac{n-1}{2}}\Psi_{\lambda,k-1}(x).
    \end{eqnarray*}
Moreover, $\Psi_{\lambda,k}$ is an eigenfunction for the magnetic Laplacian (\ref{magneticLaplacian}) with eigenvalue $k+\frac{n}{2}$.
\item For each $\Psi_k \in \mathcal{F}_k^{+-}$ such that $\Psi_k$ is a Clifford-Hermite function of degree $k$ (see also Theorem \ref{SeriesExpansionSolutions}), the function $$P_{\lambda,k}(x)=\pi^{\frac{n}{4}}e^{\frac{|x|^2}{2}}\Psi_{\lambda,k}(x),$$ with $\Psi_{\lambda,k}(x)=\exp\left( \frac{\lambda}{n}D\right)\Psi_k(x),$ gives a generating function in terms of Clifford-Hermite polynomials and corresponds to a solution of the eigenvalue problem
    $$ \Delta f(x)-\frac{2\lambda}{n}Xf(x)=2Ef(x)-2kf(x).$$
\item For each $\Psi_k \in \mathcal{F}_k^{+-}$ such that $\Psi_k$ is a Clifford-Hermite function of degree $k$ (see also Theorem \ref{SeriesExpansionSolutions}), the function $$P_{\lambda,k}(x)=\pi^{\frac{n}{4}}\mathbb{P}\left(e^{\frac{|x|^2}{2}}\Psi_{\lambda,k}(x)\right),$$ with $\Psi_{\lambda,k}(x)=\exp\left( \frac{\lambda}{n}D\right)\Psi_k(x),$ corresponds to a solution of the coupled system of equations:
\begin{eqnarray*}
\left\{\begin{array}{lll}
(D-\lambda I)f(x)=-\frac{2\lambda}{n}~\Gamma \left(f(x)\right) \\ \ \\ 2E(f(x))-2kf(x)=-\frac{2\lambda}{n}X f(x)
\end{array}\right..
\end{eqnarray*}
Moreover, $P_{\lambda,k}(x)$ is harmonic, i.e. $P_{\lambda,k}\in \ker \Delta$.
\end{enumerate}
\end{theorem}

\begin{remark}
This approach comprises Xu's approach \cite{Xu2} in the case when
$\Gamma f(x)=0$ (e.g. $f$ is a radial function). In particular, a
series representation for the null solutions of $D-\lambda I$ can
be computed by using a generating function that encodes the
eigenfunctions of the Landau operator
${\mathcal{H}}_{\lambda}=\frac{1}{2}\left((D-\lambda
I)(D+\lambda I)-X^2-\frac{2\lambda}{n}X\right)$.

Physically speaking, Xu's approach describes an electromagnetic
field without the action of orbital electromagnetic sources (cf.
\cite{BBG76,GHJ02}).
\end{remark}

The above given results also provide us with some intriguing
relations in terms of Perelomov coherent states for nilpotent Lie
groups (cf.\ \cite{CnopsKisil99}, Subsection 1.3) that yield
discrete series representations of the group $SU(1,1)$ being
isomorphic to the symplectic group $\mbox{Sp}(2,\BR)$. We will
leave this issue for a future research topic (for further details
see e.g.~\cite{Perelomov86}, Chapter 14).

In conclusion, the results obtained in this section correspond to
the extension of De Gosson and Luef's approach
\cite{DeGossonLuef09}, i.e. the link between standard Weyl
calculus and Landau--Weyl calculus was obtained in the presence of
$\mathfrak{osp}(1|2)-$symmetries, our recent results should also
lead to important investigations in the study of the regularity
and hypo-ellipticity of the solutions to Schr\"odinger equations
in the context of the theory of modulation spaces.

\ack

D.~Constales was supported by project BOF/GOA 01GA0405 of Ghent University and through the  Long Term Structural Methusalem Funding by the Flemish Government. N. Faustino was
supported by {\it Funda\c c\~ao para a Ci\^encia e a Tecnologia} under the fellowship
SFRH/BPD/63521/2009 and the project PTDC/MAT/114394/2009.

The authors would like to thank Prof.\ V.V.~Kisil for pointing us to various historical aspects about Segal-Bargmann spaces and
the Heisenberg group mentioning the references \cite{Segal63,Howe80,Kisil93} after having read a preprint version of the article. We also wish to express
our gratitude to the anonymous referees for having given valuable advices to important technical aspects. This allowed us to reorganize
the presentation of the results in a more concise way.

\appendix

\section{Technical lemmata of Section \ref{HarmonicOscillator}}\label{HarmonicOscillatorAppendix}

\subsection{Proof of Lemma \ref{RodriguesFormulae}}

\proof
Recall that $
X_1,X_2,\ldots,X_n,\partial_{X_1}, \partial_{X_2}, \ldots, \partial_{X_n}
$
are the canonical generators of the Heisenberg-Weyl algebra $\mathfrak{h}_n$ and $X^2=-\sum_{k=1}^n X_k^2 $.
Then we have
\begin{eqnarray*}
\left[\partial_{X_j},\frac{1}{2} X^2\right]= -\sum_{k=1}^n\,\left[\partial_{X_j},\frac{X_k^2}{2}\right]= -\sum_{k=1}^n\,\delta_{jk}\,X_k=-X_j,
\end{eqnarray*} and analogously $\left[\partial_{X_j},-\frac{1}{2} X^2\right]=\sum_{k=1}^n\,\left[\partial_{X_j},\frac{X_k^2}{2}\right]=X_j.$

Induction on $t \in \BN$ allows us to establish that $\left[\partial_{X_j},\left(\pm\frac{1}{2} X^2\right)^t\right]=\mp t X_j\left(\pm\frac{1}{2} X^2\right)^{t-1},$ and hence,
\begin{eqnarray*}
\begin{array}{lll}
\left[\partial_{X_j},\exp\left( \pm \frac{1}{2} X^2\right)\right]&=&\sum_{t=0}^\infty \frac{1}{t!}\left[\partial_{X_j},\left(\pm\frac{1}{2} X^2\right)^t\right] \\
 &=& \sum_{t=0}^\infty \frac{\mp X_j}{(t-1)!}  \left(\pm\frac{1}{2} X^2\right)^{t-1}
 \\&=&\mp X_j\exp\left( \pm \frac{1}{2}X^2\right).
\end{array}
\end{eqnarray*}

By adding the terms $\pm X_j\exp\left( \pm \frac{1}{2}X^2\right)+\exp\left( \pm \frac{1}{2}X^2\right)\partial_{X_j}$ to both sides of the above relations, we obtain
\begin{eqnarray*}\label{expX2}
\begin{array}{ccc}
\left( \partial_{X_j}+X_j\right)\exp\left(\frac{1}{2} X^2\right)&=&\exp\left(\frac{1}{2} X^2\right)\partial_{X_j} \\\left( \partial_{X_j}-X_j\right)\exp\left( -\frac{1}{2} X^2\right)&=&\exp\left( -\frac{1}{2} X^2\right)\partial_{X_j}.
\end{array}
\end{eqnarray*}

Finally, by multiplying on both sides of the above mentioned relations from the right with the expressions   $\frac{1}{\sqrt{2}}\exp\left(-\frac{1}{2} X^2\right)$ and $-\frac{1}{\sqrt{2}}\exp\left(\frac{1}{2} X^2\right)$, respectively, and taking into account the definitions of $A_j^\pm$ (see \ref{WeylHeisenbergFock}), we arrive at
\begin{eqnarray}
\begin{array}{ccc}
A_j^-&=&\frac{1}{\sqrt{2}}\exp\left(\frac{1}{2} X^2\right)\partial_{X_j}\exp\left(-\frac{1}{2} X^2\right)\\
A_j^+ &=& -\frac{1}{\sqrt{2}}\exp\left( -\frac{1}{2} X^2\right)\partial_{X_j}\exp\left(\frac{1}{2} X^2\right).
\end{array}
\end{eqnarray}

Using linearity arguments, the statements for the operators $D^+$ and $D^-$ are then immediate when using their coordinate expressions.
\qed

\subsection{Proof of Lemma \ref{powersDpm}}

\proof
We will use mathematical induction to prove (\ref{DminusDplusk}). For $k=1$, we proceed as follows:

We can split $[D^-,D^+]$ according to the definition as
\begin{eqnarray*}
[D^-,D^+]=\frac{1}{2}[X+D,X-D]=-[X,D].
\end{eqnarray*} From $XD=-E-\Gamma$ and $\{ X,D\}=-2E-nI$ (Lemma \ref{xDEGamma}) and the identity $2XD=\{ X,D\}+[X,D]$ it follows that $[X,D]=-2\Gamma+nI$ and hence
\begin{eqnarray}
\label{DminusDplus1}[D^-,D^+]=2\Gamma-nI=2\left( \Gamma-\frac{n}{2}I\right).
\end{eqnarray}
For $k=2$, notice that from Lemma \ref{xDEGamma} we have $\{\Gamma,X\}=(n-1)X$ and $\{\Gamma, D\}=(n-1)D$,
and hence, $\{ \Gamma, D^+\}=(n-1)D^+.$

Thus, relying on $[D^-,D^+]=2\Gamma-nI$, we obtain
\begin{eqnarray}
\label{DminusDplus2}
\begin{array}{lll}
D^-(D^+)^2&=&(2\Gamma-nI)D^+ + D^+(D^-D^+) \\
&=& (2\Gamma-nI)D^+ + D^+(2\Gamma-nI)+(D^+)^2D^-\\
&=&2\left(\left\{ \Gamma,D^+\right\}-nD^+\right)+(D^+)^2D^- \\
&=&-2D^+ + (D^+)^2 D^-.
\end{array}
\end{eqnarray}
Next, we assume that (\ref{DminusDplusk}) holds for $k \in \BN$, i.e.
\begin{eqnarray*}
D^-(D^+)^k=\left\{
\begin{array}{lll}
-2j (D^+)^{2j-1}+(D^+)^{2j}D^-, & \mbox{if}~k=2j \\ \ \\
2 (D^+)^{2j}\left( \Gamma-\left(\frac{n}{2}+j\right)I\right)+(D^+)^{2j+1}D^-, & \mbox{if}~k=2j+1
\end{array}
\right.
\end{eqnarray*}
Hence, the induction assumption together with the relations (\ref{DminusDplus1}) and (\ref{DminusDplus2}) lead to
\begin{eqnarray*}
D^-(D^+)^{2j+2}&=&\left(D^-(D^+)^{2j+1}\right)D^+\\
&=&\left( 2 (D^+)^{2j}\left( \Gamma-\left(\frac{n}{2}+j\right)I\right)+(D^+)^{2j+1}D^- \right)D^+ \\
&=& 2 (D^+)^{2j}\left( \Gamma-\left(\frac{n}{2}+j\right)I\right)D^++(D^+)^{2j+1}(D^-D^+)\\
&=& 2 (D^+)^{2j}\left( \Gamma-\left(\frac{n}{2}+j\right)I\right)D^++(D^+)^{2j+1}\left(2\left( \Gamma-\frac{n}{2}I \right)+D^+D^-\right)\\
&=& 2 (D^+)^{2j} \{ \Gamma,D^+\}-2(n+j+1)(D^+)^{2j+1}+(D^+)^{2j+2}D^-\\
&=& -(2j+2)(D^+)^{2j+1} + (D^+)^{2j+2}D^-
\end{eqnarray*}
\begin{eqnarray*}
D^-(D^+)^{2j+3}&=&\left(D^-(D^+)^{2j+2}\right)D^+\\
&=&\left( -(2j+2)(D^+)^{2j+1} + (D^+)^{2j+2}D^- \right)D^+ \\
&=&-(2j+2)(D^+)^{2j+2}+(D^+)^{2j+2}(D^-D^+)\\
&=&-2(j+1)(D^+)^{2j+2}+2(D^+)^{2j+2}\left(\Gamma-\frac{n}{2}I+\frac{1}{2}D^+D^-\right)\\
&=&-2(D^+)^{2j+2}\left( \Gamma-\left(\frac{n}{2}+j+1\right)I\right)+(D^+)^{2j+3}D^-.
\end{eqnarray*}
This proves (\ref{DminusDplusk}).
\qed

\subsection{Proof of Lemma \ref{H0Intertwining}}

\proof
Recall that from relations (\ref{LieSuperalgebraRelations}) the operators $P^+=-\frac{1}{4}\Delta$, $P^-=\frac{1}{2} X^2$ and $Q=\frac{1}{2}\left(E+\frac{n}{2}I\right)$ satisfy
\begin{eqnarray*}
\begin{array}{lll}
[Q,P^+]=P^+, & [Q,P^-]=-P^-, & [P^-,P^+]=Q.
\end{array}
\end{eqnarray*}

{\bf Proof of relation (\ref{Statement1}):}
\\
From $[P^-,P^+]=Q$ we obtain the relation $\left[ -\frac{1}{2}\Delta, \frac{1}{2} X^2 \right]= E+\frac{n}{2}I$. By  an induction argument on $k \in \BN$ we obtain
\begin{eqnarray*}
-\frac{1}{2}\Delta \left(\frac{1}{2} X^2\right)^k=\\
=\left(E+\frac{n}{2}I +\frac{1}{2} X^2\left(-\frac{1}{2}\Delta\right)\right)\left(\frac{1}{2} X^2\right)^{k-1} \\
=\left(E+\frac{n}{2}I \right)\left(\frac{1}{2} X^2\right)^{k-1}+\left(E+\frac{n}{2}I \right)+\frac{1}{2} X^2\left(-\frac{1}{2}\Delta\right)\left(\frac{1}{2} X^2\right)^{k-2} \\
=\left(E+\frac{n}{2}I \right)\left(\frac{1}{2} X^2\right)^{k-1}+\left(\frac{1}{2} X^2\right)^{2} \left(-\frac{1}{2}\Delta\right)^2\left(\frac{1}{2} X^2\right)^{k-2}  \\
= \ldots \\
=k\left(E+\frac{n}{2}I \right)\left(\frac{1}{2} X^2\right)^{k-1}+\left(\frac{1}{2} X^2\right)^{k} \left(-\frac{1}{2}\Delta\right)^2,
\end{eqnarray*}
or equivalently, $\left[-\frac{1}{2}\Delta,\left(\frac{1}{2} X^2\right)^k\right]=k\left(E+\frac{n}{2}I \right)\left(\frac{1}{2} X^2\right)^{k-1}$. This leads to
$$\left[-\frac{1}{2}\Delta,\exp\left(\frac{1}{2} X^2\right)\right]=\left(E+\frac{n}{2}I \right)\exp\left(\frac{1}{2} X^2\right).$$

On the other hand, applying the same order of ideas, by induction over $k \in \BN$ one can show that
$\left[E+\frac{n}{2}I,\left(\frac{1}{2} X^2\right)^k\right]=k\left(\frac{1}{2} X^2\right)^{k}=k\frac{1}{2} X^2\left(\frac{1}{2} X^2\right)^{k-1}$.
Hence,
$$ \left[E+\frac{n}{2}I,\exp\left(\frac{1}{2} X^2\right)\right]=\frac{1}{2} X^2\exp\left(\frac{1}{2} X^2\right).$$

Combining the above mentioned two relations with each other, then the commuting relation
$$ \left[-\frac{1}{2}\Delta+ E+\frac{n}{2}I,\exp\left(\frac{1}{2} X^2\right)\right]
=\left(\frac{1}{2} X^2+E+\frac{n}{2}I\right)\exp\left(\frac{1}{2} X^2\right)$$
follows from linearity. After straightforward simplifications we immediately get
$$\left(-\frac{1}{2}\Delta-\frac{1}{2} X^2\right)\exp\left(\frac{1}{2} X^2\right)=\exp\left(\frac{1}{2} X^2\right)\left( -\frac{1}{2}\Delta+ E+\frac{n}{2}I \right),$$
that is, $\mathcal{H}_0\exp\left(\frac{1}{2} X^2\right)=\exp\left(\frac{1}{2} X^2\right)\mathcal{J}_0$.
\\

{\bf Proof of relation (\ref{Statement2})}
\\

For the proof of (\ref{Statement2}), we observe that the relation $[Q,P^-]=-P^-$ leads to $\left[ E+\frac{n}{2}I ,-\frac{1}{4}\Delta\right]=\frac{1}{2}\Delta$. Following the same order of ideas that we applied in the proof of (\ref{Statement1}), we get $$\left[ E+\frac{n}{2}I ,\exp\left(-\frac{1}{4}\Delta\right) \right]=\frac{1}{2}\Delta\exp\left(-\frac{1}{4}\Delta\right).$$ This is equivalent to $\mathcal{J}_0\exp\left(-\frac{1}{4}\Delta\right)=\exp\left(-\frac{1}{4}\Delta\right)\left( E+\frac{n}{2}I \right)$.
\qed

\section{Technical lemmata of Section \ref{MaxwellEquations}} \label{MaxwellEquationsAppendix}

\subsection{Proof of Lemma \ref{DisplacedIntertwine}}

\proof
Recall that from $XD=-E-\Gamma$ and $\{ X,D\}=-2E-nI$ (Lemma \ref{xDEGamma}) and the identity $2XD=\{ X,D\}+[X,D]$, it follows that $[X,D]=-2\Gamma+nI$. Hence,
$$
\left[D,\frac{\lambda}{n}(D-X)\right]=\frac{\lambda}{n}[X,D]=-\frac{2\lambda}{n}\Gamma+\lambda I=\left[X,\frac{\lambda}{n}(D-X)\right]
$$
Applying mathematical induction over $k \in \BN_0$ results into
\begin{eqnarray*}
\left[D,\left(\frac{\lambda}{n}( X-D)\right)^k\right]&=&k\left(-\frac{2\lambda}{n}\Gamma+\lambda I \right) \left(\frac{\lambda}{n}(D-X)\right)^{k-1} \\
\left[X,\left(\frac{\lambda}{n}(
X-D)\right)^k\right]&=&k\left(-\frac{2\lambda}{n}\Gamma+\lambda
I \right) \left(\frac{\lambda}{n}(D-X)\right)^{k-1}.
\end{eqnarray*}
Therefore, from the series expansion of
$\exp\left(\frac{\lambda}{n}\left(D-X\right)\right)$, we obtain that
\begin{eqnarray*}
\left[D,\exp\left(\frac{\lambda}{n}\left(D-X\right)\right)\right]&=&\left(-\frac{2\lambda}{n}\Gamma+\lambda I \right) \exp\left(\frac{\lambda}{n}\left(D-X\right)\right) \\
\left[X,\exp\left(\frac{\lambda}{n}\left(D-X\right)\right)\right]&=&\left(-\frac{2\lambda}{n}\Gamma+\lambda
I \right) \exp\left(\frac{\lambda}{n}\left(D-X\right)\right).
\end{eqnarray*}
After applying straightforward algebraic manipulations, we arrive at
\begin{eqnarray*}
\exp\left(\frac{\lambda}{n}\left(D-X\right)\right)D=\left(D+\frac{2\lambda}{n}\Gamma-\lambda I\right)\exp\left(\frac{\lambda}{n}\left(D-X\right)\right) \\
\exp\left(\frac{\lambda}{n}\left(D-X\right)\right)X=\left(X+\frac{2\lambda}{n}\Gamma-\lambda I\right)\exp\left(\frac{\lambda}{n}\left(D-X\right)\right).
\end{eqnarray*}
\qed

\subsection{Proof of Lemma \ref{expLambaDminusX}}

\proof
 From the definition of $\exp\left(\frac{\lambda}{n}D\right)$ we know that we can split the series described above in the way

$$\exp\left(\frac{\lambda}{n}D\right)=\cosh\left(\frac{\lambda}{n}D\right)+\sinh\left(\frac{\lambda}{n}D\right),$$
with
\begin{eqnarray*}
\begin{array}{lll}
\cosh\left(\frac{\lambda}{n}D\right)=\sum_{k=0}^\infty \frac{\lambda^{2k}}{n^{2k}(2k)!}(-\Delta)^k,  &
\sinh\left(\frac{\lambda}{n}D\right)=\sum_{k=0}^\infty \frac{\lambda^{2k+1}}{n^{2k+1}(2k+1)!}D(-\Delta)^k.
\end{array}
\end{eqnarray*} From (\ref{LieSuperalgebraRelations}) we know that the operators
$P^-=-\frac{1}{4}\Delta$, $P^+=\frac{1}{2} X^2$ and
$Q=\frac{1}{2}\left( E+\frac{n}{2}I\right)$ (canonical generators
of $\mathfrak{sl}_2(\BR)$) satisfy
\begin{eqnarray*}
\begin{array}{lll}
[P^-,P^+]=Q, & [Q,P^-]=-P^-, & [Q,P^+]=P^+.
\end{array}
\end{eqnarray*}

This leads to $\left[-\frac{1}{2}\Delta,-\frac{1}{2} X^2\right]=-\left(E+\frac{n}{2}I\right)$. Therefore,
\begin{eqnarray}
\left[-\frac{1}{2}\Delta,\exp\left( -\frac{1}{2} X^2\right)\right]=-\left(E+\frac{n}{2}I\right)\exp\left( -\frac{1}{2} X^2\right)
\end{eqnarray}
follows straightforwardly from induction arguments and from the formal series expansion of $\exp\left( \frac{1}{2} X^2\right)$.

This equality is equivalent to
$\left(-\frac{1}{2}\Delta+E+\frac{n}{2}I\right)\exp\left( -\frac{1}{2} X^2\right)=\exp\left( -\frac{1}{2} X^2\right)\left( -\frac{\Delta }{2} \right)$.
So, we may further derive that
\begin{eqnarray*}
\exp\left( -\frac{1}{2} X^2\right)\cosh\left(\frac{\lambda}{n}D\right)\exp\left( \frac{1}{2} X^2\right)
=\\
=\sum_{k=0}^\infty \frac{\lambda^{2k}}{n^{2k}(2k)!}\exp\left( -\frac{1}{2} X^2\right)(-\Delta)^k\exp\left( \frac{1}{2} X^2\right) \\
=\sum_{k=0}^\infty \frac{\lambda^{2k}}{n^{2k}(2k)!}2^k\left(-\frac{1}{2}\Delta+E+\frac{n}{2}I\right)^k.
\end{eqnarray*}

For the computation of $\exp\left( -\frac{1}{2} X^2\right)\sinh\left(\frac{\lambda}{n}D\right)\exp\left( \frac{1}{2} X^2\right)$
we rely on the relation \begin{center}$(D-X)\exp\left(-\frac{1}{2} X^2\right)=\exp\left(-\frac{1}{2} X^2\right)D$\end{center}
 which follows from (\ref{DexpX2}).

Standard computations yield
\begin{eqnarray*}
\exp\left(-\frac{1}{2} X^2\right)\sinh\left(\frac{\lambda}{n}D\right)\exp\left(\frac{1}{2} X^2\right)= \\
=\sum_{k=0}^\infty \frac{\lambda^{2k+1}}{n^{2k+1}(2k+1)!}\exp\left(-\frac{1}{2} X^2\right)D(-\Delta)^k\exp\left( \frac{1}{2} X^2\right)\\
=\sum_{k=0}^\infty \frac{\lambda^{2k+1}}{n^{2k+1}(2k+1)!}2^k(D-X)\left(-\frac{1}{2}\Delta+E+\frac{n}{2}I\right)^k.
 \end{eqnarray*}
In view of (\ref{HamiltonianJ0}) the term $2^k\left(-\frac{1}{2}\Delta+E+\frac{n}{2}I\right)^k$ appearing on the summands is equal to $(2\mathcal{J}_0)^k$ for each $k$, so we can complete the proof of
Proposition \ref{expLambaDminusX} by applying linearity arguments.
\qed

\subsection{Proof of Lemma \ref{DpmIntertwining}}

\proof
When we follow the same lines of  the proof of Lemma
\ref{DisplacedIntertwine}, then we obtain for
$D^-=\frac{1}{\sqrt{2}}(X-D)$ and $D^+=\frac{1}{\sqrt{2}}(X+D)$ the following:
\begin{eqnarray*}
\left[ D^\mp,\exp\left(\frac{\lambda}{n}D\right)\right]=\frac{\lambda}{n\sqrt{2}}~[X,D]\exp\left(\frac{\lambda}{n}D\right) =  \left(\frac{\lambda}{\sqrt{2}} I-\frac{\sqrt{2} \lambda}{n}\Gamma\right)\exp\left(\frac{\lambda}{n}D\right).
\end{eqnarray*}

A rearrangement the terms in the identity presented above leads to
\begin{eqnarray*}
\exp\left(\frac{\lambda}{n}D\right) D^+=\left( D^+-\frac{\lambda}{\sqrt{2}} I+\frac{\sqrt{2} \lambda}{n}\Gamma \right)\exp\left(\frac{\lambda}{n}D\right) \\ \exp\left(\frac{\lambda}{n}D\right) D^-=\left( D^- -\frac{\lambda}{\sqrt{2}} I+\frac{\sqrt{2} \lambda}{n}\Gamma \right)\exp\left(\frac{\lambda}{n}D\right).
\end{eqnarray*}
Finally, the equations (\ref{DpmLambda}) follow after
straightforward algebraic manipulations. \qed

\subsection{Proof of Lemma \ref{magneticLaplacianSplitting}}

\proof

In view of the definition we can split $\left\{ X- (D+{\lambda}I)+\frac{2 \lambda}{n}\Gamma ,X+ (D-{\lambda}I)+\frac{2 \lambda}{n}\Gamma \right\}$ as
\begin{eqnarray}
\label{magneticLapl}
\begin{array}{lll}\left\{ X- (D+{\lambda}I)+\frac{2 \lambda}{n}\Gamma ,X+ (D-{\lambda}I)+\frac{2 \lambda}{n}\Gamma \right\}=\\=-\left\{D+\lambda I,D-\lambda I \right\}-\left\{D+\lambda I,X+\frac{2 \lambda}{n}\Gamma \right\}
+\\+\left\{X+\frac{2 \lambda}{n}\Gamma,D-\lambda I \right\}+\left\{ X+\frac{2 \lambda}{n}\Gamma,X+\frac{2 \lambda}{n}\Gamma\right\}.
\end{array}
\end{eqnarray}
The terms $-\left\{D+\lambda I,D-\lambda I \right\}$ and
$-\left\{D+\lambda I,X+\frac{2 \lambda}{n}\Gamma \right\}+\left\{X+\frac{2 \lambda}{n}\Gamma,D-\lambda I \right\}$ are equal to $2(\Delta+\lambda^2 I)$ and
$-4\lambda\left( X+\frac{2 \lambda}{n}\Gamma\right)$, respectively, while
 \begin{eqnarray*}
\left\{ X+\frac{2 \lambda}{n}\Gamma,X+\frac{2 \lambda}{n}\Gamma\right\} &=&2\left( X+\frac{2 \lambda}{n}\Gamma\right)^2\\
 &=&2 \left( X^2+\left\{X,\frac{2 \lambda}{n}\Gamma \right\}+\left(\frac{2 \lambda}{n}\Gamma\right)^2\right)\\
 &=&2 X^2+4\lambda\left(1-\frac{1}{n} \right)X+2\left(\frac{2 \lambda}{n}\Gamma\right)^2
 \end{eqnarray*}
follows from $\{ \Gamma,X\}=(n-1)X$ (Lemma \ref{xDEGamma}) and from the decomposition 

$(S+T)^{2}=S^2 +\{ S,T\}+T^2$ for each $S$ and $T$.

Thus, rearranging all the above expressions, equation (\ref{magneticLapl}) is equivalent to
$$ 2\left( \Delta+\lambda^2I\right)+2X^2-\frac{4\lambda}{n} X -4\lambda \left(I-\frac{1}{n}\Gamma\right)\frac{2\lambda}{n}\Gamma
=-4\mathcal{H}_\lambda,$$
which is equivalent to $\left\{ \frac{1}{\sqrt{2}}\left(X- (D+{\lambda}I)+\frac{2 \lambda}{n}\Gamma\right) ,\frac{1}{\sqrt{2}}\left(X+ (D-{\lambda}I)+\frac{2 \lambda}{n}\Gamma\right) \right\}=-2\mathcal{H}_\lambda$.
This completes our proof.
\qed

\section{Table}\label{Table}

\begin{center}
\begin{tabular}{|c|l|}
  \hline
  Symbol &  Notation/Explanation \\
  \hline \\
  $\{ S,T\}$ \& $[S,T]$ & $\{ S,T\}=ST+TS$ ; $[S,T]=ST-TS$  \\ & \\
  $\exp(S)$ &  $\exp(S)=\sum_{j=0}^\infty \frac{1}{k!}S^k$  \\ & \\
  $\BR_{0,n}$ & Clifford algebra of signature $(0,n)$   \\ & \\
  $\mbox{End}(\mathcal{T})$ & algebra of endomorphisms of $\mathcal{T}$  \\ & \\
  $\mathcal{P}$ & $\mathcal{P}=\BR[\underline{x}]\otimes \BR_{0,n}$  \\ & \\
  $\mathcal{P}_j$ & $\mathcal{P}_j=\left\{f \in \mathcal{P}~:~f(t\underline{x})=t^j f(\underline{x}), ~\forall~ t \in \BR, ~\forall ~\underline{x} \in \BR^n \right\}$  \\ & \\
  $L_2(\BR^n;\BR_{0,n})$ & $L_2(\BR^n;\BR_{0,n}):=L_2(\BR^n)\otimes \BR_{0,n}$  \\ & \\
  $\mathcal{S}(\BR^n)$ \& $\mathcal{S}(\BR^n;\BR_{0,n})$ &  $\mathcal{S}(\BR^n)$ Schwartz space over $\BR^n$; $\mathcal{S}(\BR^n;\BR_{0,n}):=\mathcal{S}(\BR^n)\otimes \BR_{0,n}$ \\ & \\
  $D$ \& $\Delta$ & $D=\sum_{j=1}^n \xi_j \partial_{X_j}$;~$\Delta=\sum_{j=1}^n \partial_{X_j}^2$ \\ & \\
  $X$ \& $X^2$ & $X=\sum_{j=1}^n \xi_j X_j$;~$X^2=-\sum_{j=1}^n X_j^2$ \\ & \\
  $E$ \& $\Gamma$ & $E=\sum_{j=1}^n  X_j\partial_{X_j}$; $\Gamma=\sum_{j=1}^n  \xi_j \xi_k \left( X_j\partial_{X_k}-X_k\partial_{X_j}\right)$\\ & \\
  $D^\pm$ & $D^\pm=\frac{1}{\sqrt{2}}\left( X \mp D \right)$ \\ & \\
  $\mathcal{H}_0$ \& $\mathcal{J}_0$ & $\mathcal{H}_0=-\frac{1}{2}\Delta+\frac{1}{2}|x|^2I$;~ $\mathcal{J}_0=-\frac{1}{2}\Delta+E+\frac{n}{2}I$ \\ & \\
$P_j^\Delta(x)$ & $P_j^\Delta(x)=\exp\left( -\frac{\Delta}{4}\right)P_j(x)$, with $P_j\in \mathcal{P}_j$ \\ & \\
$\phi(x)$ \& $\Psi(x)$ & $\Psi(x)=\phi(x)=\pi^{-\frac{n}{4}}e^{-\frac{|x|^2}{2}}$ \\ & \\
$c_k$ & $c_k=\left(k+\frac{n-1}{2} \right)\left(k-1+\frac{n-1}{2} \right)\ldots \left(1+\frac{n-1}{2} \right)\frac{n-1}{2}$ \\ & \\
$\Psi_k(x)$ \& $\psi_k(x)$ & $\Psi_k(x)=\frac{1}{\sqrt{c_k}}(D^+)^k(\Psi(x))$;~$\psi_k(x)=\pi^{\frac{n}{4}}e^{\frac{|x|^2}{2}}\Psi_k(x)$\\ & \\
 $\mathbb{P}$ \& $\mathbb{Q}$ & $\mathbb{P}=I+\frac{1}{2k+n-2}XD$;~$\mathbb{Q}=-\frac{1}{2k+n-2}X D$ \\ 
  \hline
\end{tabular}
\end{center}

\end{document}